\documentclass[12pt]{iopart}

\usepackage{iopams}  
\usepackage{graphicx}

\newcommand{\id}{\mathbb{I}}
\newcommand{\Z}[0]{{\mathbb{Z}}}
\def\C{{\mathbb{C}}}
\def\Z{{\mathbb{Z}}}

\def\N{{\mathbb{N}}}

\newcommand{\cH}{{\cal H}}

\newcommand{\cS}{{\cal S}}

\newcommand{\ket}[1]{|#1\rangle}
\newcommand{\bra}[1]{\langle #1|}

\begin{document}

\title[Eulerian Hamiltonian simulation]
{Hamiltonian quantum simulation with bounded-strength controls}

\author{Adam D. Bookatz$^1$, Pawel Wocjan$^{2}$, and Lorenza Viola$^{3}$}

\address{$^1$Center for Theoretical Physics, Massachusetts Institute
of Technology, Cambridge, Massachusetts 02139, USA; \texttt{bookatz@mit.edu}}

\address{$^2$Department of Electrical Engineering 
and Computer Science, University of Central Florida, Orlando, Florida 32816, USA; \texttt{wocjan@eecs.ucf.edu}}

\address{$^3$Department of Physics and Astronomy, Dartmouth College,
6127 Wilder Laboratory, Hanover, New Hampshire 03755, USA; \texttt{lorenza.viola@dartmouth.edu}}

\date{\today}

\begin{abstract}
We propose dynamical control schemes for Hamiltonian simulation in
many-body quantum systems that avoid instantaneous control operations 
and rely solely on realistic {\em bounded-strength control Hamiltonians}.
Each simulation protocol consists of periodic repetitions of a basic control
block, constructed as a suitable modification of an ``Eulerian
decoupling cycle,'' that would otherwise implement a
trivial (zero) target Hamiltonian.  For an open quantum system coupled
to an uncontrollable environment, our approach may be employed to
engineer an effective evolution that simulates a target Hamiltonian on
the system, while suppressing unwanted decoherence to the leading
order.  We present illustrative applications to both closed- and
open-system simulation settings, with emphasis on {\em simulation of 
non-local (two-body) Hamiltonians using only local (one-body) controls}.  
In particular, we provide simulation schemes applicable to 
Heisenberg-coupled spin chains exposed to general linear decoherence, 
and show how to simulate Kitaev's honeycomb lattice Hamiltonian starting
from Ising-coupled qubits, as potentially relevant to the dynamical 
generation of a topologically protected quantum memory.  Additional 
implications for quantum information processing are discussed.
\end{abstract}
\qquad\qquad\quad\ \, {{\sc mit-ctp} 4504}

\section{Introduction}

The ability to accurately engineer the Hamiltonian of complex quantum 
systems is both a fundamental control task and a prerequisite for quantum 
simulation, as originally envisioned by Feynman \cite{Feynman,Lloyd,Nori}. 
The basic idea underlying Hamiltonian simulation is to use an available quantum 
system, together with available (classical or quantum) control resources, to emulate 
the dynamical evolution that would have occurred under a different, desired Hamiltonian 
not directly accessible to implementation \cite{Remark1}. From a control-theory 
standpoint, the simplest setting is provided by {\em open-loop Hamiltonian 
engineering in the time domain} \cite{Schirmer,Domenico}, whereby coherent control 
over the system of interest is achieved solely based on suitably designed 
time-dependent modulation (most commonly sequences of control pulses), 
without access to ancillary quantum resources and/or measurement and 
feedback. While open-loop Hamiltonian engineering techniques have their origin and 
a long tradition in nuclear magnetic resonance (NMR) \cite{WHH:68,EBW:87}, the 
underlying physical principles of ``coherent averaging'' have recently found 
widespread use in the context of quantum information processing (QIP), leading 
in particular to dynamical symmetrization and dynamical decoupling (DD) schemes 
for control and decoherence suppression in open quantum systems 
\cite{QECbook,bangbang,violaDecoupling,violaControl,Zanardi:00,Viola2002}. 

As applications for quantum simulators continue to emerge across a vast array of 
problems in physics and chemistry, and implementations become closer to experimental 
reality \cite{Nori,Blatt,Britton}, it becomes imperative to expand the repertoire of 
available Hamiltonian simulation procedures, while scrutinizing the validity of the 
relevant control assumptions. With a few exceptions (notably, the use of so-called 
``perturbation theory gadgets'' \cite{Terhal2008}), open-loop  
Hamiltonian simulation schemes have largely relied thus far on the ability to 
implement sequences of effectively instantaneous, ``bang-bang'' (BB) control pulses 
\cite{WJB:02,DNBT:2002,finite,BCLLLPV:2002,univControl,Squeezing,Becker2013,Loss2013}. 
While this is a convenient and often reasonable first approximation, instantaneous pulses 
necessarily involve unbounded control amplitude and/or power, something which is out of 
reach for many control devices of interest and is fundamentally unphysical. 
In the context of DD, a general approach for achieving (to at least the leading order) 
the same dynamical symmetrization as in the BB limit was proposed in \cite{VK:02}, 
based on the idea of continuously applying bounded-strength control Hamiltonians according 
to an Eulerian cycle, so-called {\em Eulerian DD} (EDD).  From a Hamiltonian engineering 
perspective, EDD protocols translate directly into bounded-strength simulation schemes for 
{\em specific} effective Hamiltonians -- most commonly, the trivial (zero) Hamiltonian 
in the case of ``non-selective averaging'' for quantum memory (or ``time-suspension'' 
in NMR terminology). More recently, EDD has also served as the starting point for 
bounded-strength {\em gate simulation} schemes in the presence of decoherence, 
so-called {\em dynamically corrected gates} (DCGs) for universal quantum 
computation \cite{KV09a,KV09b,KLV10,KBV12}.

In this work, we show that the approach of Eulerian control can be further 
systematically exploited to construct {\em bounded-strength Hamiltonian simulation schemes} 
for a broad class of target evolutions on both closed and open (finite-dimensional) quantum systems.  
Our techniques are device-independent and broadly applicable, thus substantially expanding the 
control toolbox for programming complex Hamiltonians into existing or near-term quantum 
simulators subject to realistic control assumptions.  
 
The content is organized as follows.  We begin in Sect.~II by introducing the 
appropriate control-theoretic framework and by reviewing the basic principles 
underlying open-loop simulation via average Hamiltonian theory, along with its 
application to Hamiltonian simulation in the BB setting.  
Sect.~III is devoted to constructing and analyzing simulation schemes that employ 
bounded-strength controls: while Sec. III.A reviews required 
background material on EDD, Sec. III.B introduces our new Eulerian simulation 
protocols for a generic closed quantum system. In Sec. III.C we separately 
address the important problem of Hamiltonian simulation in the presence of 
slowly-correlated (non-Markovian) decoherence, identifying conditions under 
which a desired Hamiltonian may be enacted on the target system while 
simultaneously decoupling the latter from its environment, and making further 
contact with DCG protocols. Sect.~IV presents a number of illustrative applications 
of our general simulation schemes in interacting multi-qubit networks.  
In particular, we provide explicit protocols to simulate a large family of 
two-body Hamiltonians in Heisenberg-coupled spin systems additionally 
exposed to depolarization or dephasing, as well as to achieve Kitaev's honeycomb 
lattice Hamiltonian starting from Ising-coupled qubits.  In all cases, 
only {\em local} (single-qubit, possibly collective) control Hamiltonians 
with bounded strength are employed.  A brief summary and outlook conclude
in Sec. V.

\section{Principles of Hamiltonian simulation}

\subsection{Control-theoretic framework}
\label{sec:framework}

We consider a quantum system $\cS$, with associated Hilbert space
$\cH$, whose evolution is described by a time-independent Hamiltonian 
$H$.  As mentioned, \emph{Hamiltonian simulation} is the task of making
$\cS$ evolve under some other time-independent target Hamiltonian,
say, $\tilde{H}$. Without loss of generality, both the input and the target 
Hamiltonians may be taken to be traceless.  Two related scenarios are worth
distinguishing for QIP purposes:

$\bullet$ {\bf Closed-system simulation}, in which case $\cS$ coincides
with the quantum system of interest, $S$ (also referred to as the ``target'' 
henceforth), which undergoes purely {\em unitary} (coherent) dynamics;

$\bullet$ {\bf Open-system simulation}, in which case $\cS$ is
a bipartite system on $\cH\equiv \cH_S \otimes \cH_B$, where 
$B$ represents an uncontrollable environment (also referred to as bath henceforth), 
and the reduced dynamics of the target system $S$ is {\em non-unitary} in general.
 
In both cases, we shall assume the target system $S$ to be a network
of interacting qudits, hence $\cH_S \simeq ({\mathbb C}^d)^{\otimes
n}$, for finite $d$ and $n$. 
In the general open-system scenario, the
joint Hamiltonian on $\cH$ may be expressed in the following form,
\begin{equation}
H = H_S \otimes \id_B + \id_S \otimes H_B + \sum_\alpha S_\alpha
\otimes B_\alpha ,
\label{openHam}
\end{equation} 
where the operators $H_S$ ($H_B)$ and $S_\alpha$ ($B_\alpha$) act on
$\cH_S$ ($\cH_B$) respectively, and all the bath operators are assumed to
be norm-bounded, but otherwise unspecified (potentially unknown). The
closed-system setting is recovered from Eq.~(\ref{openHam}) 
in the limit $S_\alpha\equiv 0$.
Likewise, we may express the target Hamiltonian $\tilde{H}$ in a
similar form, with two simulation tasks being of special relevance:
$\tilde{S}_\alpha \equiv 0$, in which case the objective is to realize
a desired system Hamiltonian $\tilde{H}_S$ while dynamically
decoupling $S$ from its bath $B$, thereby suppressing unwanted
decoherence \cite{violaDecoupling}; or, more generally, $H_S \mapsto 
\tilde{H}_S$ {\em and} $S_\alpha \mapsto \tilde{S}_\alpha$, where the simulated, 
dynamically symmetrized error generators $\tilde{S}_\alpha$ may allow for 
decoherence-free subspaces or subsystems to exist \cite{Zanardi:00,dygen}.

Formally, the dynamics is modified by an
open-loop controller acting on the target system according to
\begin{equation}
H \mapsto H(t) = H + H_c(t), \hspace{8mm} H_c(t) \equiv 
\sum_u h_u(t) = \sum_u f_u(t) X_u,
\label{controlledHam}
\end{equation} 
where the operators $\{ X_u =X_u^\dagger \}$ and the (real)
functions $\{f_u(t)\}$ represent the available control Hamiltonians and
the corresponding, generally time-dependent, control inputs respectively.  
Clearly, if the Hamiltonian $(\tilde{H}-H)$ is contained
in the admissible control set, the corresponding control problem is
trivial and the desired time-evolution, 
$$\tilde U(t) = e^{-i \tilde H t},\quad t \geq 0,$$
\noindent 
can be exactly simulated continuously in time.
However, this level of control need not be available in settings of
interest, including open quantum systems where control actions are
necessarily restricted to the target system $S$ alone, $H_c(t)\equiv H_c(t) \otimes
\id_B$ in Eq.~(\ref{controlledHam}).  Following the general idea of
``analog'' quantum simulation \cite{Nori}, we shall assume in
what follows a \emph{restricted} set of control Hamiltonians (in a
sense to be made more precise later) and focus on the task of
{\em approximately} simulating the desired time evolution $\tilde U(t)$ 
at a {\em final time} $t=\tilde{T}_f$, or more generally, \emph{stroboscopically}
in time, that is, at instants $t=\tilde t_M$, where
\[ \tilde t_M = M \tilde T, \qquad M\in\N, \]
and $\tilde T$ is a {fixed} minimum time interval.  Choosing $\tilde T$
sufficiently small allows in principle any desired accuracy in the
approximation to be met, with the limit $\tilde T \rightarrow 0$
formally recovering the continuous limit.  

Specifically, let $U(t)$ and $U_c(t)$ denote the unitary propagators associated to the total 
and the control Hamiltonians in Eq.~(\ref{controlledHam}), respectively:
\begin{eqnarray}
U(t) &=&{\cal T}\exp\left\{ -i \int_0^t
[H+H_c(\tau)]\,d\tau\right\}\,, \label{propagator}\\
U_c(t) &=&{\cal T}\exp\left\{ -i \int_0^t
H_c(\tau) \,d\tau\right\}\,, 
\label{Uc}
\end{eqnarray}
where we have set $\hbar=1$ and ${\cal T}$ indicates time-ordering, as
usual. Then, for a given pair $(H,\tilde{H})$, we shall provide
sufficient conditions for $\tilde{H}$ to be ``reachable'' from $H$ and, if
so, devise a \emph{cyclic} control procedure such that the resulting controlled propagator
\begin{equation}
\label{eq:introgoal}
U(t_M) \approx  \tilde U(\tilde t_M), \qquad t_M = M T_c\,, \qquad M\in\N ,
\end{equation}
where $T_c$ is the {cycle time} of the controller, 
that is, $U_c(t+T_c) =U_c(t)$. In general, we shall allow for $T_c$ 
to differ from $\tilde T$, corresponding to an overall scale factor in the 
simulated time, as it will become apparent later. If, for a {\em fixed} input 
Hamiltonian $H$, {\em arbitrary} target Hamiltonians are reachable for 
given control resources, the simulation scheme is referred to 
as \emph{universal}.  In this case, complete controllability must be 
ensured by the tunable Hamiltonians $X_u$ in conjunction with the ``drift'' 
$H_S$ \cite{Domenico}. In contrast, we shall be especially interested in 
situations where control over $S$ is more limited.

Similar to DD protocols, Hamiltonian simulation
protocols are most easily constructed and analyzed by 
effecting a transformation to the ``toggling'' frame associated
to $U_c(t)$ in Eq.~(\ref{Uc}) \cite{EBW:87,violaDecoupling,Viola2002}.  
That is, evolution in the toggling frame is generated by the time-dependent, 
control-modulated Hamiltonian 
\begin{equation}
H'(t)= U_c^\dagger(t) H U_c(t), 
\label{eq:tHam}
\end{equation} 
with the corresponding toggling-frame propagator $U'(t)$ being related 
to the physical propagator in Eq.~(\ref{propagator}) by
$U(t)= U_c(t)U'(t)$.
Since the control propagator is cyclic and $H$ is time-independent,
it follows that $U(t_M) = U'(t_M)$ and, furthermore, $H'(t)$ acquires 
the periodicity of the controller, $U'(t_M) = [U'(T_c)]^M$. Thus, the 
stroboscopic controlled dynamics of the system is determined by
\begin{equation}
\label{eq:U=U'}
U(t_M) = [U'(T_c)]^M\,. 
\end{equation}
Average Hamiltonian theory \cite{EBW:87,Haeberlen76} may then be invoked
to associate an effective \emph{time-independent} Hamiltonian $\bar{H}$ to
the evolution in the toggling-frame: 
\begin{equation}
\label{eq:effective_U}
U(T_c)=U'(T_c) \equiv \exp(-i \bar{H} T_c) \,,
\end{equation}
where $\bar{H}$ is determined by the Magnus expansion \cite{Magnus:54}, 
$\bar{H}=\bar{H}^{(0)}+\bar{H}^{(1)}+\bar{H}^{(2)}+\ldots$ Explicitly,
the leading-order term, determining evolution over a cycle up to the
first order in time, is given by
\begin{equation}
\label{eq:approx}
\bar{H}^{(0)}
=\frac{1}{T_c}
\int_0^{T_c}H'(\tau)d\tau
=\frac{1}{T_c}
\int_0^{T_c} U_c^\dagger(\tau) H U_c(\tau)\, d\tau\,,
\end{equation}
with (absolute) convergence being ensured as long as 
$t \Vert H\Vert < \pi$ \cite{BCOR: 2009}.  Subject to convergence condition, 
higher-order corrections for evolution over time $t$ can also be upper-bounded 
by (see Lemma 4 in \cite{Kaveh2008})
\begin{equation} 
\Big\Vert \sum_{\ell=\kappa}^\infty t \bar{H}^{(\ell)} \Big\Vert \leq 
c_\kappa [ \,(t \Vert H \Vert)^{\kappa +1}\,], \quad c_\kappa = O(1).
\label{bound}
\end{equation}

Ideally, one would like to achieve $\bar{H} T_c = \tilde{H} \tilde{T}$, so 
that equality would hold in Eq.~(\ref{eq:introgoal}) for all $M \in {\mathbb N}$.  
In what follow, we shall primarily focus on achieving \emph{first-order simulation} 
instead, by engineering the control propagator $U_c(t)$ in such a way that
\begin{eqnarray}
\label{eq:introgoal_long}
\bar{H} T_c \approx \bar{H}^{(0)} T_c = \tilde{H} \tilde{T}, 
\end{eqnarray}
whereby, using Eq.~(\ref{bound}) with $\kappa =1$, 
\begin{eqnarray} 
U(t_M) = e^{-i  \bar{H} M T_c} = e^{-i \bar{H}^{(0)} M T_c} + 
O [ ( M T_c \Vert H \Vert )^2 ] 
\approx \tilde{U}(\tilde{t}_M). 
\label{eq:introgoal_long2}
\end{eqnarray}
In general, the accuracy of the approximation in Eq.~(\ref{eq:introgoal_long})
improves as the ``fast control limit'', $T_c\rightarrow 0$, is
approached.  Physically, this corresponds to requiring that the shortest 
control time scale (pulse separation) involved in the control sequence be 
sufficiently small relative to the shortest correlation time of the dynamics 
induced by $H$ \cite{Haeberlen76,KavehLimits}. 
While the problem of constructing general high-order Hamiltonian simulation 
schemes is of separate interest, we stress that {\em second-order simulation} 
can be readily achieved, in principle, by ensuring that $U_c(t)$ is time-symmetric, 
namely, $U_c(t)=U_c(T_c-t)$ for $t\in [0,T_c]$. Since all odd-order Magnus corrections vanish in this case \cite{Haeberlen76}, it follows (by using again Eq.~(\ref{bound}), with $\kappa =2$), that $\bar H T_c = \tilde H \tilde T + O[(\Vert H \Vert T_c)^3]$, correspondingly boosting the
accuracy of the simulation.

\subsection{Hamiltonian simulation with bang-bang controls} 
\label{sec:bang-bang}

BB Hamiltonian simulation provides the simplest control setting for achieving 
the intended objective, given in Eq.~(\ref{eq:introgoal}).  Two main assumptions 
are involved: 
(i) First, we must be able to express the target Hamiltonian $\tilde{H}$ as 
\begin{equation}
\label{eq:bang-bang-scheme}
\tilde{H}=\sum_{j=1}^N w_{j} U_{j}^\dagger H U_{j} 
\,, \hspace*{8mm} W\equiv \sum_j w_j\,,
\end{equation}
where  $\{ U_j \}$ are unitary operators on $S$ and the $\{ w_j\}$ non-negative 
real numbers (not all zero).
(ii) Second, the available control resources include a discrete set of instantaneous pulses $\{ P_j \}$ on $S$, whose application results in a piecewise-constant control propagator $U_c(t)$  over $[0, T_c]$, with corresponding toggling-frame 
propagators $\{ U_j \}$, $U_j \equiv \prod_{k=1}^j P_k$, $U_1=\id_S$ \cite{QECbook,Viola2002}. Assumptions (i)-(ii) together allow for the time-average in Eq.~(\ref{eq:approx}) to be mapped to a convex (positive-weighted) sum.  Eq.~(\ref{eq:bang-bang-scheme}) may be interpreted as a \emph{sufficient} condition for the target Hamiltonian $\tilde{H}$ to be reachable from $H$ given open-loop unitary control on $S$ alone.  Reachable Hamiltonians must thus be at least as ``disordered'' as the input one in the sense of majorization \cite{BCLLLPV:2002,Viola2002}. 

Specifically, Eq.~(\ref{eq:bang-bang-scheme}) leads naturally to the following BB simulation scheme. Given simulation weights $\{w_j\}$, define the following simulation intervals and timing pattern:
\begin{equation}
\label{BBtimings}
\tau_j \equiv w_j \tilde T , 
\quad \quad t_j \equiv \sum_{k=1}^j \tau_{k},\;t_0= 0\,,\; t_N \equiv T_c = W\tilde{T}\,.
\end{equation}
A piecewise-constant control propagator for the basic simulation block to be repeated 
may then be constructed as follows: 
\begin{equation}
\label{eq:control}
U_c^{\rm{BB}} (t_{j-1}+ \theta) = U_{j}\,,\quad\theta\in [0,\tau_{j}] ,\, \hspace{8mm}
j=1,\ldots,N \,.
\end{equation}
By using Eq. ~(\ref{eq:approx}), it is immediate to verify that 
\begin{eqnarray}
\bar{H}^{(0)} =
\frac{1}{T_c} \sum_{j=1}^N \tau_{j} U_{j}^\dagger H U_{j} 
= \frac{\tilde T}{T_c}\tilde{H}\,, 
\end{eqnarray}
resulting in the desired controlled evolution, Eqs.~(\ref{eq:introgoal_long})-(\ref{eq:introgoal_long2}), 
provided that the convergence conditions 
for first-order simulation under $H$ are obeyed.  Since, in practice, technological limitations always constrain the cycle duration to a {\em finite} minimum value $T_c>0$, such conditions ultimately determine the maximum simulated time 
$\tilde{t}_M$ up to which evolution under $\tilde{H}$ may be reliably simulated 
using the physical Hamiltonian $H$. 

In analogy with BB DD schemes, realizing the prescription of 
Eq.~(\ref{eq:control}) requires to discontinuously change the control propagator 
from $U_{j}$ to $U_{j+1}=(U_{j+1} U_{j}^\dagger) U_{j}$, via an instantaneous BB 
pulse $U_{j+1} U_{j}^\dagger$ at the $j$th endpoint $t_j$. 
As a result, despite its conceptual simplicity, BB simulation is unrealistic 
whenever large control amplitudes are not an option, and the evolution induced 
by $H$ {\em during} the application of a control pulse must be considered from the 
outset. This demands redesigning the basic control block in such a way that the 
actions of $H$ and $H_c(t)$ are simultaneously accounted for.

\section{Hamiltonian simulation with bounded controls}

\subsection{Eulerian simulation of the trivial Hamiltonian} 
\label{sec:edd}

The key to overcome the disadvantages of BB Hamiltonian simulation is 
to ensure that the control propagator varies smoothly (continuously) 
in time during each control cycle.  We achieve this goal by relying on 
\emph{Eulerian control design} \cite{VK:02}. To introduce the necessary 
group-theoretical background, we begin by revisiting how, for the special 
case of a target identity evolution (that is, $\tilde{H}\equiv 0$, also 
corresponding to a ``noop'' gate, in terms of the end-time simulated propagator), 
EDD can be naturally interpreted as a bounded-strength simulation scheme. 

In the Eulerian approach, the available control resources include a discrete set of 
unitary operations on $S$, say, $\{ U_\gamma\}$, $\gamma = 1, \ldots, L$, which are 
realized  over a finite time interval $\Delta$ through application of bounded-strength 
control Hamiltonians $\{ h_\gamma (t) \},$ $\gamma =1, \ldots, L$, with 
$| h_\gamma (t)| \leq h_{\rm{max}}< \infty$.  That is, 
\begin{equation}
\label{eq:control1}
U_\gamma  \equiv u_\gamma(\Delta)\,, \quad 
u_\gamma (\delta) = {\cal T} \exp \Big\{-i\int_0^\delta h_\gamma (\tau) d\tau \Big\} \,. 
\end{equation}
Note that the choice of the control Hamiltonians $h_\gamma (t)$ is not unique, 
which allows for implementation flexibility.  The unitaries $\{ U_\gamma\}$  are identified with the image of a generating set of a finite group under a faithful, unitary, projective representation $\rho$~\cite{VK:02}.  That is, let ${\mathcal G}\equiv \{ g \}$ be a finite group of order $|{\mathcal G}|$, such that each element may be written as an ordered product of elements in a generating set 
$\Gamma \equiv \{\gamma \}$ of order $|\Gamma|=L$, $g \mapsto \rho(g)\equiv U_g$ be the representation map~\cite{Remark2}, and $G\equiv \{ U_g\}$.  The \emph{Cayley graph} $C({\cal G}, \Gamma)$ of ${\cal G}$ relative to $\Gamma$ can be thought of as pictorially representing all elements of ${\cal G}$ as strings of generators in $\Gamma$.  Each vertex represents a group element and a vertex $g$ is connected to another vertex $g'$ by a directed edge ``colored'' (labeled) with generator $\gamma$ if and only if $g'=\gamma g$.  The number of edges in $C({\cal G}, \Gamma)$ is thus equal to $N \equiv |\Gamma||{\mathcal G}|$.  Because a Cayley graph is regular, it always has an \emph{Eulerian cycle} that visits each edge exactly once and starts (and ends) 
on the same vertex \cite{Bollobas:98,GR:01}.  Let us denote with ${\mathcal C}\equiv (\gamma_1,\ldots, \gamma_N)$ the ordered list of generators defining an Eulerian cycle on $C({\cal G}, \Gamma)$ which, without loss of generality, starts (and ends) at the identity element of ${\mathcal G}$.  

Once a control Hamiltonian for implementing each generator as in Eq.~(\ref{eq:control1}) is chosen, an EDD protocol is constructed by assigning a cycle time as $T_c \equiv N \Delta$ and by applying the control Hamiltonians $h_\gamma(t)$ sequentially in time, following the order determined by the Eulerian cycle ${\cal C}$. Thus, 
\begin{equation}
U_c^{\rm{EDD}}(t_j)= U_{\gamma_j} U_c^{\rm{EDD}}(t_{j-1})\,,\quad j=1,\ldots, N\,, 
\label{eq:edd}
\end{equation}
where $U_{\gamma_j}$ is the image of the generator labeling the $j$th edge in ${\mathcal C}$.  As established in \cite{VK:02}, the lowest-order average Hamiltonian associated to the above EDD cycle has the form $\bar{H}^{(0)} \equiv \Pi_{\cal G}[F_\Gamma(H)]$, where 
for any operator $A$ acting on ${\cH}_S$, the map 
\begin{equation}
\label{eq:groupAverage}
\Pi_{\cal G}(A) = \frac{1}{|{\cal G}|} \sum_{g\in {\cal G}} U_g^\dagger A U_g 
\end{equation}
projects onto the centralizer of ${\mathcal G}$ 
(i.e., $\Pi_{\cal G}(A)$ commutes with all $U_g \in G$), and 
\begin{equation}
\label{eq:F}
F_\Gamma (H)= \frac{1}{|\Gamma|}
\sum_{\gamma\in \Gamma} \frac{1}{\Delta} 
\int_0^\Delta u_\gamma(\tau)^\dagger H u_\gamma (\tau) d\tau
\end{equation}
implements an average of $H$ over both the control interval and the group generators. 
Accordingly, bounded-strength simulation of $\tilde{H}=0$ is achieved provided that the following DD condition is obeyed:
\begin{equation} 
\label{eq:DDF}
\Pi_{\cal G}\big[ F_\Gamma (H)\big] = 0\,.
\end{equation}
By Schur's lemma, this is automatically ensured if the group representation acts irreducibly on ${\cal H}_S$. Formally, the BB limit may be recovered by letting $F_\Gamma (A)\equiv A$ for all $A$  \cite{VK:02}, reflecting the ability to directly implement all the group elements (with no overhead, as if $|\Gamma|=1$) and corresponding to uniform simulation weights $w_j =1/|{\cal G}|$.

\subsection{Eulerian simulation protocols beyond {noop}: Construction} 
\label{sec:eulerian}

\begin{figure}[t] 
\begin{center}
\includegraphics[width=15.5cm]{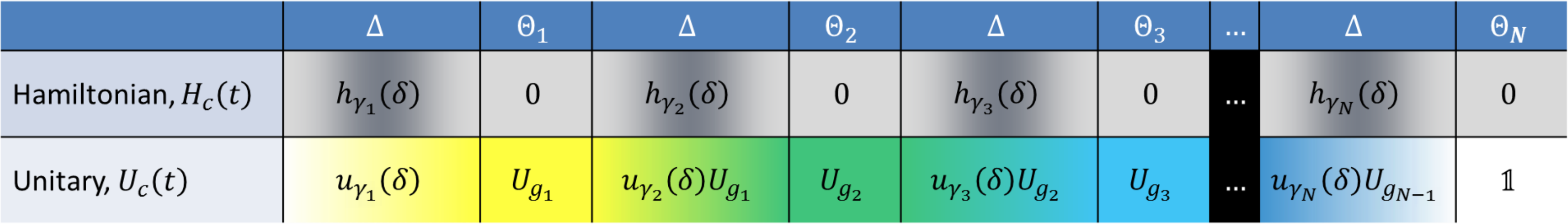}
\caption{\label{fig:protocols} Schematics of an Eulerian simulation protocol. 
The basic control block consists of $N$ time intervals, each involving 
a ``ramping-up'' sub-interval of fixed duration $\Delta$, during which $H_c(t) 
\ne 0$, followed by a ``coasting'' (free evolution) period of variable duration 
$\Theta_k$, Eq.~(\ref{Thetaj}), during which no control is applied.  
During the $j$th ramping-up sub-interval we apply $h_{\gamma_j}$, i.e., the control 
Hamiltonian that realizes the generator $\gamma_j$, smoothly changing the control propagator from $U_{g_{j-1}}$ to $U_{g_j}$. In this way, the control protocol corresponding to Eqs.~(\ref{eq:implement})-(\ref{eq:wait}) is implemented.  By construction, a standard EDD cycle with $\tilde H =0$ is recovered by letting $\Theta_k\rightarrow 0$ for all $k$, while in the limit $\Delta \rightarrow 0$ standard BB simulation of $\tilde{H}$ is implemented. }
\end{center}
\end{figure}

We show how the Eulerian cycle method can be extended to bounded-strength simulation of a non-trivial class of target Hamiltonians.  We assume that $\tilde{H}$ may be expressed as a convex unitary mixture of the group representatives $U_g$, 
\begin{equation}
\label{eq:bbSimulation}
\tilde{H}=\sum_{g \in {\cal G}} w_g U_g^\dagger H U_g 
\,, \quad \quad w_g \geq 0, \; W= \sum_g w_g \,.
\end{equation}
We construct the desired control protocol starting from an Eulerian cycle ${\cal C}=(\gamma_1,\ldots, \gamma_N)$ on $C({\cal G}, \Gamma)$.  Specifically, the idea is to append to each of the $N$ control slots that define an EDD scheme a free-evolution (or ``coasting'') period of suitable duration, in such a way that the net simulated Hamiltonian is modified from $\tilde{H}=0$ to $\tilde{H} \ne 0$ as given in Eq.~(\ref{eq:bbSimulation}). A pictorial representation of the basic control block is given in Fig. \ref{fig:protocols}.  As in Eq.~(\ref{eq:control1}), let $\Delta$ denote the minimum time duration required to implement each generator, hence, to smoothly change the control propagator from a value $U_{g}$ to $U_{g'}$ along the cycle.  While such ``ramping up'' control intervals have all the same length, each ``coasting'' interval is designed to keep the control propagator constant at $U_{g'}$ for a duration determined by the corresponding weight $w_{g'}$. Since the control is switched off during coasting, continuity of the overall control Hamiltonian $H_c(t)$ may be ensured, if desired, by requiring that 
\begin{equation}  
h_\gamma (0)=0 = h_\gamma (\Delta)\,, \quad \gamma=1,\ldots, L, 
\label{continuity}
\end{equation}
in addition to the bounded-strength constraint. 

An Eulerian simulation protocol may be formally specified as follows.  As before, let 
the $j$th time interval be denoted as $[t_{j-1},t_j]$, $j=1,\ldots, N$, 
with $t_0 = 0$ and $t_N$ defining the cycle time $T_c$.  For each $j$, let $\tau_{g_j} \equiv w_{g_j} \tilde T$ as in the BB case.  The duration of the $j$th coasting period is then assigned as 
\begin{equation}
\Theta_j \equiv \frac{\tau_{g_j}}{|\Gamma|}\,,
\label{Thetaj}
\end{equation} 
resulting in the following timing pattern $\{t_j\}$ [compare to Eq.~(\ref{BBtimings})]:
\begin{equation}
\label{EUtimings}
t_j = \sum_{k=1}^j (\Delta + \Theta_k) = 
j\Delta + \frac{1}{|\Gamma|} \sum_{k=1}^j \tau_{g_k}\,,\quad 
t_N \equiv T_c = N\Delta + W \tilde T\,.
\end{equation}
As the expression for the cycle times makes it clear, the resulting protocol may be equivalently interpreted in two ways: starting from an EDD cycle, corresponding to $N\Delta$ and $\tilde{H}= 0$, we introduce the coasting periods to allow for non-trivial simulated dynamics to emerge; or, starting from a BB simulation scheme for $\tilde{H}$, corresponding to $W \tilde T$, we introduce the ramping-up periods to allow for control Hamiltonians to be smoothly switched over $\Delta$.  Either way, bounded-strength protocols imply a time-overhead $N\Delta$ relative to the BB case, 
recovering the BB limit as $\Delta \rightarrow 0$ as expected.  Explicitly, the control propagator for Eulerian simulation has the form:
\begin{eqnarray}
U_c^{\rm{EUS}}\big( t_{j-1} + \delta \big) = u_{\gamma_j}(\delta) U_{g_{j-1}}\quad &
\quad \rm{for \ } \delta\in [0,\Delta] \,,
\label{eq:implement} \\ 
U_c^{\rm{EUS}} \big(t_{j-1} + \Delta + \theta \big) = 
U_{g_j}\,\quad &\quad \rm{for \ } \theta\in [0,\Theta_j] \,.
\label{eq:wait}
\end{eqnarray}

The resulting first-order Hamiltonian $\bar{H}^{(0)}$ under Eulerian
simulation is derived by evaluating the time-average in
Eq.~(\ref{eq:approx}) with the control propagator given by Eqs. 
(\ref{eq:implement})-(\ref{eq:wait}).  Direct calculation along the lines of 
\cite{VK:02} yields: 
\begin{eqnarray*}
\bar{H}^{(0)} & = &  \frac{1}{T_c} 
\sum_{j=1}^N
\bigg[
\int_{\delta=0}^\Delta U_c(t_{j-1}+\delta)^\dagger H U_c(t_{j-1}+\delta) d\delta 
\nonumber \\ 
& \mbox{} & \hspace*{11mm}+
\int_{\theta=0}^{\Theta_j} U_c(t_{j-1}+\Delta+\theta)^\dagger H U_c(t_{j-1}+\Delta+\theta) d\theta \bigg] \nonumber \\ 
& = &
\frac{1}{T_c} \sum_{j=1}^N 
\bigg[
\int_{\delta=0}^\Delta U_{g_{j-1}}^\dagger u_{\gamma_j}(\delta)^\dagger H u_{\gamma_j}(\delta) U_{g_{j-1}} d\delta +
\int_{\theta=0}^{\Theta_j} U_{g_j}^\dagger H U_{g_j} d\theta
\bigg] \nonumber \\ 
&=&
\frac{1}{T_c} \sum_{g\in {\cal G}}
\bigg[ U_g^\dagger \bigg(
\sum_{\gamma \in \Gamma} \int_{\delta=0}^\Delta u_\gamma(\delta)^\dagger H u_\gamma(\delta) d\delta \bigg)
U_g + |\Gamma|\frac{\tau_g}{|\Gamma|} U_g^\dagger H U_g
\bigg] , 
\label{eq:fourth}
\end{eqnarray*}
where the last equality follows from two basic properties of Eulerian cycles: 
firstly, the list $\{g_0,g_1,\ldots,g_{N-1}\}$ (and also $\{g_1,g_2,\ldots,g_N\}$) 
of the vertices that are being visited contains each element $g\in {\cal G}$ precisely 
$|\Gamma|$ times; secondly, in traversing the Cayley graph, each group element $g$ 
is left exactly once by a $\gamma$-labeled edge for each generator $\gamma \in \Gamma$.  
Thus, by recalling the definitions given in Eqs.~(\ref{eq:groupAverage}) and 
(\ref{eq:F}), we finally obtain
\begin{eqnarray} 
\bar{H}^{(0)}  = \frac{N \Delta}{T_c} \Pi_{\cal G}\big[F_\Gamma (H)\big] + 
\frac{\tilde T}{T_c} \sum_{g\in {\cal G}} w_g U_g^\dagger H U_g = 
\frac{\tilde T}{T_c} \tilde H \, ,
\label{eq:fifth}
\end{eqnarray}
which indeed achieves the intended first-order simulation goal, Eqs.~(\ref{eq:introgoal_long})-(\ref{eq:introgoal_long2}), as long as convergence holds and 
the DD condition of Eq. (\ref{eq:DDF}) is obeyed. 

The simulation accuracy may be improved by
symmetrizing $U^{\rm{EUS}}_c(t)$ in time.  In analogy to symmetrized EDD protocols \cite{QECbook}, this can be easily accomplished by running the protocol and
then suitably running it again in reverse. Specifically, let the duration of the coasting 
interval be changed as $\Theta_j \mapsto\Theta_j/2$. Run the protocol as described above until time $t = N\Delta + \frac{1}{2}W\tilde T$. 
Then, from time $t = N\Delta + \frac{1}{2}W\tilde T$ until time 
$t=T_c = 2N\Delta + W\tilde T$, modify Eqs.~(\ref{eq:implement})-(\ref{eq:wait}) 
as follows:
\begin{eqnarray*}
U_c^{\rm{EUS}}\big[ T_c - (t_{j-1} +\Delta) + \delta \big] = u_{\gamma_j}(\Delta -
\delta) U_{g_{j-1}} \ &\quad \rm{for \ } \delta\in [0,\Delta] \,,\\ 
U_c^{\rm{EUS}} 
\big[ T_c - (t_{j-1} +\Delta +\Theta_j) + \theta \big] = U_{g_j} \ &\quad \rm{for\ 
} \theta\in [0,\Theta_j] \,,
\end{eqnarray*}
for $j=N,\ldots,1$. Provided that one is able to implement
$u_{\gamma_j}(\Delta - \delta)$, we again obtain
\[ \bar H^{(0)} = 2\frac{N\Delta}{T_c} \Pi_{\cal G}\big[F_\Gamma (H)\big] + \frac{\tilde
T}{T_c} \sum_{g\in {\cal G}} w_g U_g^\dagger H U_g\, , \]
while satisfying $U_c(t)=U_c(T_c-t)$ for $t\in
[0,T_c]$, hence ensuring that $\bar H^{(1)} = 0$.

\subsection{Eulerian simulation while decoupling from an environment}
\label{sec:enviro}

The ability to implement a desired Hamiltonian on the target system $S$, 
while switching off (at least to the leading order) the coupling to an uncontrollable environment $B$, is highly relevant to realistic applications.   That is, with reference to Eq.~(\ref{openHam}), the objective is now to \emph{simultaneously} achieve $\tilde{H}_S\equiv H_{\rm{target}}, \tilde{S}_\alpha \equiv 0$, by unitary control operations acting on $S$ alone.
Because the first-order Magnus term $\bar H^{(0)}$ is additive [recall 
Eq. ~(\ref{eq:approx})], it is appropriate to treat each summand of $H$
individually, leading to a relevant average Hamiltonian of the form 
$$\bar H^{(0)}= \bar{H}_S \otimes \id_B + \sum_\alpha \bar{S}_\alpha \otimes B_\alpha +\id_S \otimes H_B \,,\;$$
where for a generic operator on $\cH_S$ we let  
$$\bar {A} \equiv \frac{1}{T_c}\int_0^{T_c} U_c^\dagger(\tau) A U_c(\tau)\,
d\tau\,. $$
We can then apply the analysis of Sec. \ref{sec:eulerian} to the internal system Hamiltonian ($\bar{H}_S$) and each error generator ($\bar{S}_\alpha$) separately, to obtain in both cases a simulated operator of the form given in Eq.~(\ref{eq:fifth}):
\begin{equation*}
\label{eq:A_bar}
\bar A = \frac{N \Delta}{T_c} \Pi_{\cal G}\big[ F_\Gamma (A)\big] + \frac{\tilde T}{T_c}
\sum_{g\in {\cal G}} w_g U_g^\dagger A U_g\,.
\end{equation*}

Since the task is to decouple $S$ from $B$ while maintaining the non-trivial evolution due to $\tilde H_S=H_{\rm{target}}$, the reachability condition of 
Eq.~(\ref{eq:bbSimulation}) must now ensure that
\begin{eqnarray}
\tilde H_S &=&\sum_{g\in {\cal G}} w_g U_g^\dagger H_S U_g \,, \qquad
\label{eq:enviro_assume_1}\\ 
0 &=& \sum_{g\in {\cal G}} w_g U_g^\dagger
S_\alpha U_g \,,\quad \forall\alpha\,. 
\label{eq:enviro_assume_2}
\end{eqnarray}
Accordingly, it is necessary to extend the DD assumption of
Eq.~(\ref{eq:DDF}) to become 
\begin{eqnarray} 
\Pi_{\cal G}\big[F_\Gamma(H_S)\big] & = & 0 \,, \qquad \quad
\label{eq:enviro_assume_3} \\
\Pi_{\cal G}\big[F_\Gamma(S_\alpha)\big] & = & 0 \,,\;\;\forall\alpha \,,\label{eq:enviro_assume_4}
\end{eqnarray}
such that $\bar A = ({\tilde T}/{T_c})  \tilde A$ holds for each of the summands 
in $H$.  Altogether we recover
\[
\bar{H}^{(0)} = \frac{\tilde T}{T_c}\tilde{H}_S\otimes\id_B  +\id_S\otimes H_B.
\]

It is interesting in this context to highlight some similarities and differences with 
DCGs \cite{KV09a}, which also use Eulerian control as their starting point and are specifically designed to achieve a desired unitary evolution on the target system while simultaneously removing decoherence to the leading \cite{KV09a,KV09b,KBV12} or, in principle, arbitrarily high order \cite{KLV10}.  By construction, the open-system simulation procedure just described \emph{does} provide a first-order DCG implementation for the target gate $Q\equiv \exp(-i \tilde{H}_S \tilde{T}_f)$: in particular, the requirement that Eqs.~(\ref{eq:enviro_assume_1})-(\ref{eq:enviro_assume_2}) be obeyed together (for the {\em same} weights $w_g$) is effectively equivalent to evading the ``no-go theorem'' for black-box DCG constructions established in \cite{KV09b}, with the coasting intervals and the resulting ``augmented'' Cayley graph playing a role similar in spirit to a (first-order) ``balance-pair'' implementation.  Despite these formal similarities, a number of differences exist between the two approaches: first, an obvious yet important difference is that DCG constructions focus directly on synthesizing a desired unitary {\em propagator}, as opposed to a desired Hamiltonian {\em generator}. Second, while the internal system Hamiltonian, $H_S$, is a crucial input in a Hamiltonian simulation problem, it is effectively treated as an unwanted error contribution in analytical DCG constructions, in which case complete controllability over the target system must be supplied by the controls alone.  Although in more general (optimal-control inspired) DCG constructions \cite{KBV12}, limited external control is assumed and $H_S$ may become essential for universality to be maintained, emphasis remains, as noted above, on end-time synthesis of a target propagator.  Finally, a main intended application of DCGs is realizing low-error {\em single- and two-qubit gates} for use within fault-tolerant quantum computing architectures, as opposed to robust Hamiltonian engineering for many-body quantum simulators which is our focus here.

\subsection{Eulerian simulation protocols: Requirements} 
\label{sec:reasonability}

Before presenting explicit illustrative applications, we summarize 
and critically assess the various requirements that should be obeyed for Eulerian simulation to achieve the intended control objective of Eq.~(\ref{eq:introgoal}) 
in a closed or, respectively, open-system setting:

\begin{enumerate}
\item \label{assume:Hamiltonian} {\em Time independence}.
Both the internal Hamiltonian $H$ and the target Hamiltonian $\tilde H$ 
are taken to be time-independent (and, without loss of generality, traceless).

\item \label{assume:group} {\em Reachability}. The target Hamiltonian 
$\tilde{H}$ must be reachable from $H$, that is, there must be a control 
group ${\mathcal G}$, with a faithful, unitary projective representation 
mapping $g \mapsto \rho(g)= U_g$, such that Eq.~(\ref{eq:bbSimulation}) holds. 
For dynamically-corrected Eulerian simulation in the presence of an environment, 
this requires, as noted, that for the \emph{same} weights $\{w_g\}$, the desired system Hamiltonian is reachable from $H_S$ while the trivial (zero) Hamiltonian is 
reachable from each error generator $S_\alpha$ separately, such that both Eqs. 
(\ref{eq:enviro_assume_1})-(\ref{eq:enviro_assume_2}) hold together. 

\item \label{assume:control} {\em Bounded control}.
For each generator $\gamma$ of the chosen control group ${\mathcal G}$, we need
access to \emph{bounded} control Hamiltonians $h_\gamma(t)$, 
such that application of $h_\gamma(t)$ 
over a time interval of duration $\Delta$ realizes the group representative 
$U_\gamma = \rho(\gamma)=u_\gamma(\Delta)$, additionally subject (if desired) to the continuity condition of Eq.~(\ref{continuity}). 

\item \label{assume:irred} {\em Decoupling conditions}.  Suitable DD conditions, 
Eq. (\ref{eq:DDF}) in a closed system or Eqs.~(\ref{eq:enviro_assume_3})-(\ref{eq:enviro_assume_4}) 
in the open-system error-corrected case, must be fulfilled, in order for undesired contributions 
to the simulated Hamiltonians to be averaged out by symmetry to the leading order. 

\item \label{assume:efficiency} {\em Time-efficiency}.  If the choice of 
${\cal G}$ is not unique for given $(H, \tilde{H})$, the smallest group should be 
chosen, in order to keep the number of intervals per cycle, $N = |{\cal G}||\Gamma|$, to a minimum.  In particular, {\em efficient} Hamiltonian simulation requires that $|{\cal G}|$ (hence also $|\Gamma|$) scales (at most) {\em polynomially} with the number of subsystems $n$. 

\end{enumerate}

The key simplification that the time-independence Assumption~(\ref{assume:Hamiltonian}) introduces into the problem is that the periodicity of the control action is directly transferred to the toggling-frame Hamiltonian of Eq.~(\ref{eq:tHam}), allowing one to simply focus on single-cycle evolution.  Although this assumption is strictly not fundamental, general time-dependent Hamiltonians may need to be dealt with on a case-by-case basis (see also \cite{RDD1,RDD2,LeaNJP}).  A situation of special practical relevance arises in this context for open systems exposed to {\em classical noise}, in which case 
$\cH_B \simeq {\mathbb C}$ and the system-bath interaction in Eq.~(\ref{openHam}) is effectively replaced by a classical, time-dependent stochastic field.  Similar to DD and DCG schemes, Eulerian simulation protocols remain applicable as long as the noise process is stationary and exhibiting correlations over sufficiently long time scales \cite{QECbook,Nave}. 
  
The reachability Assumption~(\ref{assume:group}) is a prerequisite for Eulerian Hamiltonian simulation schemes. Although BB Hamiltonian simulation need not be group-based, most BB schemes follow this design principle alike.  Assumption (\ref{assume:control}), restricting the admissible control resources to {\em physical} Hamiltonians with bounded amplitude (thus finite control durations, as opposed to instantaneous implementation of arbitrary group unitaries as in the BB case) is a basic assumption of the Eulerian control approach.  As remarked, our premise is that the available Hamiltonian control is {\em limited}, restricted to only the target system if the latter is coupled to an environment, and typically {\em non-universal} on ${\cal H}_S$; in particular, we cannot directly express $\tilde H = H + H_c$ and apply $H_c =\tilde{H}-H$, or else the problem would be trivial.  In addition to error-corrected Hamiltonian simulation in open quantum systems, scenarios of great practical interest may arise when the control Hamiltonians 
are subject to more restrictive {\em locality constraints} than the 
system and target Hamiltonians are (e.g., two-body simulation with only local controls, see also Sec. \ref{sec:example_simple}).

The required decoupling conditions in Assumption~(\ref{assume:irred}) 
are automatically obeyed if the representation $\rho$ acts irreducibly on ${\cal H}_S$.  This follows directly from Schur's lemma, together with the fact that the map $F_\Gamma$ defined in Eq. (\ref{eq:F}) is trace-preserving, and both $H_S$ and $S_\alpha$ can be taken to be traceless.  While convenient, irreducibility is not, however, a requirement.  When the representation $\rho$ is reducible, care must be taken in order to ensure that Assumption (\ref{assume:irred}) is nevertheless obeyed. It should be stressed that this is possible {\em independently} of the target Hamiltonian $\tilde H$. Therefore, if the choice (${\cal G}, \rho$) works for one Eulerian simulation scheme (whether $\rho$ is irreducible or not), then it can be used for
Eulerian simulation with any target $\tilde H$ that belongs to the reachable set from $H$, 
that is, that can satisfy Eq.~(\ref{eq:bbSimulation}).   

We close this discussion by recalling that it is always possible for a finite-dimensional 
target system $S$ to find a control group ${\cal G}$ for which both 
Assumptions~(\ref{assume:group}) and (\ref{assume:irred}) are satisfied, by resorting to the concept of a {\em transformer} \cite{univControl,Viola2002}. A transformer is a pair $({\cal G},\rho)$, where ${\cal G}$ is a finite group and $\rho: {\cal G} \rightarrow {\cal U}({\cH_S}),\, g \mapsto \rho(g)=U_g$ is a faithful, unitary, projective representation such that, for {\em any} traceless Hermitian operators $A$ and $B$ on $\cH_S$ with $A \ne 0$, one may express 
$$B=\sum_{g\in {\cal G}} w_g U_g^\dagger A U_g\,, \quad w_g \geq 0.$$
We illustrate this general idea in the simplest case of a single qubit, $\cH=\cH_S=\C^2$.  
Let $X,Y,Z$ denote the Pauli matrices and $R$ the unitary matrix defined by
\begin{equation} 
\label{eq:transformerR}
R=\frac{i-1}{2}\left(\begin{array}{cc} i & i \\ -1 & 1\end{array}\right),
\end{equation}
which corresponds to a rotation by an angle $4\pi/3$ about an axis 
$\hat{n}\equiv (1,1,1)/\sqrt{3}$. 
Direct calculation shows that $R^3=I$ and that conjugation by $R$
cyclically shifts the Pauli-matrices, i.e., $R^\dagger X R = Y,
R^\dagger Y R = Z$, and $R^\dagger Z R = X$.
Consider now the group ${\cal G}$ given by the presentation 
$${\cal G}= \langle x,y,z,r \ |\ x^2=y^2=z^2=r^3=1,xz=y,r^{-1}x r=y,r^{-1}y r=z,r^{-1}z
r=x\rangle .$$  
\noindent 
Using the defining relations of this group, its elements
can always be written as $x^a z^b r^c$, where $a,b\in\{0,1\}$ and
$c\in\{0,1,2\}$. Clearly, the assignment $\rho$ given by $x\mapsto X,
y\mapsto Y, z\mapsto Z, r\mapsto R$ yields a faithful, unitary,
irreducible representation since the Pauli matrices commute up to
phase.  It is shown in \cite{univControl} that the pair $({\cal G},\rho)$
defines a transformer in the sense given above, namely, 
any $2\times 2$ traceless matrix $B$ may be reached from any fixed 
$2\times 2$ traceless, nonzero matrix $A$, for suitable non-negative weights 
$w_g$. The irreducibility property for {\em any} transformer pair can be easily 
established by contradiction \cite{Remark3}.

A drawback of the transformer formalism 
is that general transformer groups tend to be large, 
making purely transformer-based simulation schemes inefficient. 
In practice, given the native system Hamiltonian $H_S$, the challenge is to
find a group ${\cal G}$ that grants a reasonably efficient scheme while satisfying 
Assumptions~(\ref{assume:group}) and (\ref{assume:irred}), and subject to the
ability to implement the required control operations.  As we shall see next, 
transformer-inspired ideas may still prove useful in devising simulation schemes in 
the presence, for instance, of additional symmetry conditions.  


\section{Illustrative applications}

In this section, we explicitly analyze simple yet paradigmatic Hamiltonian 
simulation tasks motivated by QIP applications.  While a number of other 
interesting examples and generalizations may be envisioned 
(as also further discussed in the Conclusions), our goal here is to give 
a concrete sense of the usefulness and versatility of our Eulerian simulation 
approach in physically realistic control settings.  In particular, we focus on 
achieving \emph{non-local Hamiltonian simulation using only bounded-strength 
local (single-qubit) control}, in both closed and open multi-qubit systems.

\subsection{Eulerian simulation in closed Heisenberg-coupled qubit networks}
\label{sec:example_simple}

Let us start from the simplest case of a system consisting of $n=2$ qubits, 
interacting via an isotropic Heisenberg Hamiltonian of the form 
\[ H = H_{\rm{iso}} = J( X\otimes X + Y\otimes Y + Z\otimes Z) 
\equiv J( X_1 X_2 + Y_1 Y_2 + Z_1 Z_2), \]
\noindent 
where $J$ has units of energy and the second equality defines an equivalent compact 
notation. We are interested in a class of target XYZ Hamiltonians of the form 
\begin{equation}
\tilde H = H_{\rm{XYZ}}= 
J_x X_1 X_2 + J_y Y_1 Y_2 + J_z Z_1 Z_2 ,\;\; J_u \in {\mathbb R}.
\label{xyz}
\end{equation}
For instance, $J_x=J_y =\pm J$, $J_z=0$ corresponds to an isotropic XX model, 
whereas if $J_x=J_y$ with $J_z \ne 0$, an XXZ interaction is obtained, the special value 
$J_z =\mp 2J$ corresponding to the important case of a dipolar 
Hamiltonian.  The construction of a simulation protocol starts from 
observing that Hamiltonians as in Eq.~(\ref{xyz}) are reachable from $H$, 
in the sense of Eq.~(\ref{eq:bbSimulation}), based on single-qubit control only. 

Specifically, let ${\mathcal G}\equiv \Z_2\times\Z_2\equiv \Z_2^2$, and let 
the representation $\rho$ map $(n,m)\in {\mathcal G}$ to $X^nZ^m\otimes\id$.  That is, 
${\mathcal G}$ is mapped to the following set of unitaries: 
\begin{equation}
\{ U_g\} = \lbrace \id\otimes\id, X\otimes\id, Y\otimes\id, Z\otimes\id \rbrace 
\equiv {G}_1 = \lbrace \id, X_1, Y_1, Z_1 \rbrace .
\label{xyzgroup}
\end{equation}
Choosing the generators of ${\mathcal G}$ to be 
$(1,0) \mapsto \gamma_{x,1} =X_1$ and $(0,1) \mapsto \gamma_{z,1} = Z_1$, 
we assume that we have access to the control Hamiltonians
\[ h_x(t) = f_x(t)X_1 \quad {\rm{ and }} \quad 
   h_z(t) = f_z(t)Z_1 \,,\]
where the control inputs $f_x(t)$ and $f_z(t)$ satisfy $f_u(0)=0=f_u(\Delta)$ 
and $\int_0^\Delta f_u(\tau)d\tau = \pi/2$, for $u=x,z$.  Recalling Eq.~(\ref{eq:control1}), 
this yields the control propagators 
\begin{eqnarray*}
u_x(\delta) & = &\cos\bigg[\int_0^\delta f_x(\tau)d\tau\bigg]
\id - i\,\sin\bigg[\int_0^\delta f_x(\tau)d\tau\bigg] X_1 \,,\\ 
u_z(\delta) & = &\cos\bigg[\int_0^\delta f_z(\tau)d\tau\bigg]
\id - i\,\sin\bigg[\int_0^\delta f_z(\tau)d\tau\bigg] Z_1 ,  
\end{eqnarray*}
with $u_x(\Delta) =X_1$ and $u_z(\Delta) =Z_1$ (up to phase), as desired.

Note that for any single-qubit Hamiltonians $A$ and $B$, averaging over the 
unitary group in Eq.~(\ref{xyzgroup}) results in the following projection 
super-operator:
\begin{equation}
\Pi_{\cal G}(A\otimes B) = \frac{1}{4} \sum_{U\in\lbrace \id,X,Y,Z\rbrace}
\hspace*{-3mm}U^\dag A U \otimes B = \frac{1}{2}\tr(A)\id\otimes B .
\label{cancel}
\end{equation}
In general, the map $F_\Gamma$ is trace-preserving and, in this case, it 
acts non-trivially only on the first qubit. Thus, $F_\Gamma$ is trace-preserving 
on the first qubit.  Since each term in $H$ is traceless in the first qubit, the decoupling condition $\Pi_{\cal G}[F_\Gamma(H)]=0$ follows directly from Eq.~(\ref{cancel}), even though the relevant representation $\rho$ is, manifestly, 
reducible. 

\begin{figure}[t] 
\begin{center}
\includegraphics[height=1.6in]{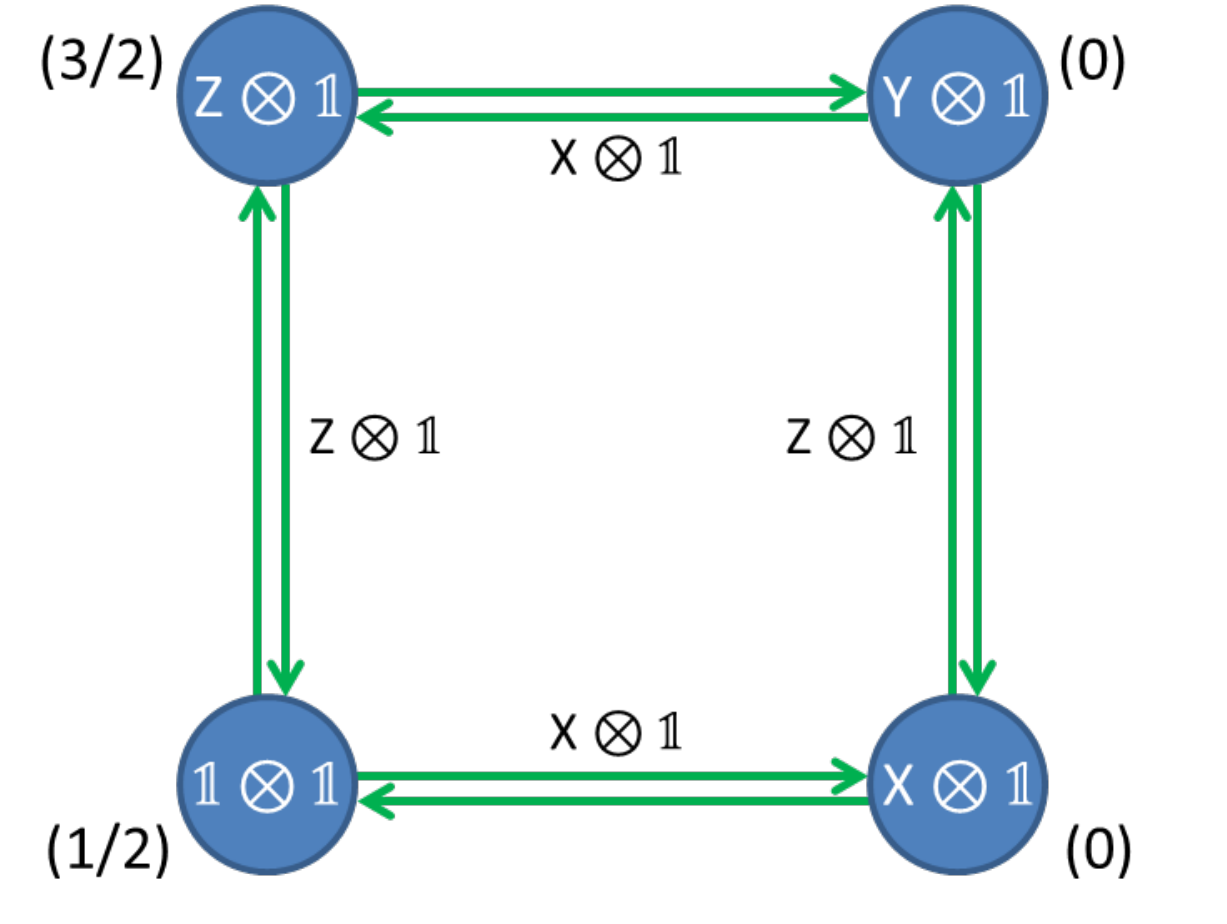}
\caption{\label{fig:cayley1}Cayley graph for the Eulerian simulation of 
the dipolar Hamiltonian in Heisenberg-coupled qubits.  Vertices are labeled by group 
elements; edges are labeled by group generators. Numbers in parentheses next to
vertices indicate the weights $w_g$ of the corresponding group
elements $g$ in Eq.~(\ref{xyz}), which is proportional to the time 
$\tau_g = w_g \tilde T$ spent at vertex $g$ during
the coasting subinterval; see also Fig.~\ref{fig:protocols}.}
\end{center}
\end{figure}

Having satisfied our main requirements for Eulerian simulation, reachability of 
XYZ Hamiltonians as in Eq.~(\ref{xyz}) is equivalent to the existence of a solution to 
the following set of conditions:
\begin{eqnarray}
J(w_{\id} + w_{X_1} - w_{Y_1} - w_{Z_1}) = J_x \,,\nonumber \\
J(w_{\id} - w_{X_1} + w_{Y_1} - w_{Z_1}) = J_y \,, \label{Js}\\
J(w_{\id} - w_{X_1} - w_{Y_1} + w_{Z_1}) = J_z \,, \nonumber 
\end{eqnarray} 
for non-negative weights $w_g$.  While infinitely many choices exist in general,
minimizing the total weight $W=\sum_g w_g$ keeps the simulation time overhead 
to a minimum.  For instance, it is easy to verify that a dipolar Hamiltonian 
of the form 
\[ \tilde H = H_{\rm{dip}} = -J\, (X_1 X_2 + Y_1 Y_2 - 2 Z_1 Z_2) \]
may be simulated with minimum time overhead by choosing weights
\[ w_{\id} = \frac{1}{2}, \quad w_{X_1} =0 =w_{Y_1} , \quad w_{Z_1} = \frac{3}{2}.
\]
The Cayley graph associated with the resulting Eulerian simulation protocol is depicted 
in Fig.~\ref{fig:cayley1}, with the explicit timing structure of the control 
block as in Fig.~\ref{fig:protocols} and $N=2\times 4=8$ control segments per block. 
It is worth observing that although the weights $w_{X_1}$ and $w_{Y_1}$
are zero in the particular case at hand, {\em all} group members of ${\cal G}$ are 
nonetheless required, and the unitaries $X_1$ and $Y_1$ still show up in the simulation 
scheme (during the ramping-up sub-intervals, as evident from Eq.~(\ref{eq:implement})).  
This is crucial to guarantee that the unwanted $F_\Gamma$ term is projected out. 

The above analysis and simulation protocols can be easily generalized to a chain of
$n$ qubits (or spins), subject to nearest-neighbor (NN) homogeneous Heisenberg 
couplings, that is, a Hamiltonian of the form
\[ H=H_{\rm{iso}}^{(\rm{NN})} = \sum_{i=1}^{n-1} J\,(
X_i X_{i+1} + Y_i Y_{i+1} + Z_i Z_{i+1})  \equiv \sum_{i=1}^{n-1} J\, 
\vec{\sigma}_i  \cdot \vec{\sigma}_{i+1},\]
where for later reference we have introduced the standard compact notation 
$\vec{\sigma}_i\equiv (X_i, Y_i, Z_i)$ and we assume for concreteness that $n$ is even. 
In this case, we need only change the unitary representation $\rho$ of $\Z_2\times\Z_2$ 
to be defined by the two generators
$(1,0) \mapsto \gamma_{x,\rm{odd}} 
=X\otimes\id\otimes X\otimes\id\otimes \cdots\otimes X\otimes\id 
\equiv X_1 X_3 \ldots X_{n-1}$ and
$(0,1) \mapsto \gamma_{z,\rm{odd}}= Z\otimes\id\otimes Z\otimes\id\otimes \cdots\otimes
Z\otimes\id \equiv Z_1 Z_3 \ldots Z_{n-1}$, resulting in the set of unitaries \cite{LeaNJP}
\[ \{ U_g\} = \lbrace \id, X_1 X_3 \ldots X_{n-1}, Y_1 Y_3 \ldots Y_{n-1}, 
Z_1 Z_3 \ldots Z_{n-1} \rbrace \equiv {G}_{\rm{odd}}, \]
Physically, the required generators $\gamma_{x, \rm{odd}}$ and $\gamma_{z,\rm{odd}}$ correspond to control Hamiltonians that are still just sums of 1-local terms, 
and that act non-trivially on odd qubits only:
\[
h_x(t) = f_x(t) (X_1+X_3 +\ldots \hspace*{-0.5mm}+X_{n-1}),\;\;
h_z(t) = f_z(t) (Z_1+Z_3 + \ldots \hspace*{-0.5mm}+ Z_{n-1}) .\]

We expect that the design of 
Eulerian simulation schemes for more general scenarios where both the input and the target 
$(H, \tilde H)$ are {\em arbitrary} two-body Hamiltonians (including, for instance, long-range couplings) will greatly benefit from the existence of combinatorial approaches for 
constructing efficient DD groups \cite{Stoll,RDD2}.   A more in-depth analysis of this topic is, however, beyond our current scope.

\subsection{Error-corrected Eulerian simulation in open Heisenberg-coupled qubit networks} 

Imagine now that the Heisenberg-coupled system $S$ considered in the previous section is 
coupled to an environment $B$, and the task is to achieve the desired XYZ Hamiltonian 
simulation while also removing \emph{arbitrary linear decoherence} to the leading order. 
The total input Hamiltonian has the form 
\begin{equation}
H = H_{\rm{iso}}^{(\rm{NN})} \otimes {\id_B} + \id_S \otimes H_B + \sum_{i=1}^n 
\vec{\sigma}_i \otimes \vec{B}_i , \quad \vec{B}_i\equiv (B_{x,i}, B_{y,i}, B_{z,i}),
\label{openchain1}
\end{equation}
where $H_B$ and $B_{u,i}$, for each $i$ and $u=x,y,z$, are operators acting on 
${\mathcal H}_B$, whose norm is sufficiently small to ensure convergence of the relevant 
Magnus series, similar to first-order DCG constructions \cite{KV09a,KV09b}.  The target Hamiltonian then reads 
\[ \tilde{H} = H_{\rm{XYZ}} \otimes \id_B +  \id_S \otimes H_B ,\]
in terms of suitable coupling-strength parameters $J_u$ as in Eq.~(\ref{xyz}).
As before, we start by analyzing the case of $n=2$ qubits in full detail.  
Our strategy to synthesize 
a dynamically corrected simulation scheme involves two stages: 
(i) We will first decouple $S$ from $B$, while leaving the system Hamiltonian 
$H_S=H_{\rm{iso}}$ unaffected; (ii) We will then apply the closed-system protocol of
Sec.~\ref{sec:example_simple} to convert $H_{\rm{iso}}$ into the target 
system Hamiltonian $\tilde H_S=H_{\rm{XYZ}}$. Once a suitable group and 
weights are identified in this way, both stages are carried out simultaneously 
in application.

A suitable DD group able to suppress general linear decoherence is provided 
by ${\cal G}_{\rm{DD}} =\Z_2 \times \Z_2$, under the $n$-fold tensor power 
representation yielding (see also \cite{KV09b}):
\[ \{ U_h \} \equiv {G}_{\rm{GL}}=
\{ \id, X^{(\rm{all})} , Y^{(\rm{all})} , Z^{(\rm{all})}\} = 
\{ \id, X_1 X_2, Y_1 Y_2, Z_1 Z_2\},  \]
generated, for instance, by the two collective generators $\gamma_{x,\rm{all}} 
= X^{(\rm{all})}=X_1 X_2$ and  $\gamma_{z,\rm{all}} = Z^{(\rm{all})}= Z_1 Z_2$.
In addition to the order of ${G}_{\rm{GL}}$ being minimal, with 
$|{G}_{\rm{GL}}| =4$ independently of $n$, step (i) above is automatically 
satisfied for the input Hamiltonian at hand, since 
\begin{equation}
[ H_{\rm{iso}} , U_h ]=0, \quad \forall U_h \in {G}_{\rm{GL}}.
\label{commute}
\end{equation}
Given a generic operator $A$ on ${\cal H}=\cH_S \otimes \cH_B$, 
we may define the superoperator $\Phi_{\rm{DD}}$ as
\[ \Phi_{\rm{DD}} (A) = \frac{1}{4} \Big( A + X^{(\rm{all})} A X^{(\rm{all})} + 
Y^{(\rm{all})} A Y^{(\rm{all})} + Z^{(\rm{all})} A Z^{(\rm{all})} \Big),
\]
corresponding to weights $\{ w_h\}$ given by $ w_{\id} = w_{X_1 X_2} =  
w_{Y_1 Y_2} =  w_{Z_1 Z_2} = {1}/{4}.$  

In step (ii), we still rely on the group $\Z_2 \times \Z_2$, but now under a 
different representation.  We choose the representation yielding the 
set ${G}_1$ of Eq.~(\ref{xyzgroup}), with the same single-qubit generators 
$\gamma_{x,1}=X_1$, $\gamma_{z,1} =Z_1$, and the corresponding weights $\{ w_{g_1}\}$ 
determined by the solution of Eqs. (\ref{Js}). 
Define the superoperator $\Phi_{1}$ to act as
\[  \Phi_{1} (A) = w_{\id} A + w_{X_1} X_1 A X_1 + w_{Y_1} Y_1 A Y_1 + 
w_{Z_1} Z_1 A Z_1 .  \]
Then the combined action of the two superoperators $\Phi_{\rm{DD}}$ and $\Phi_1$ 
yields
\begin{equation}
\Phi_1 [\Phi_{\rm{DD}} (A)] = \sum_{U_{g_1} \in {G}_1} 
\sum_{U_h \in {G}_{\rm{GL}} }
{w_{g_1}} w_h U_{g_1}^\dagger U_h^\dagger A U_h U_{g_1}  
\equiv \sum_{g \in {\cal G}} w_g U_{g}^\dagger A U_{g} , 
\label{generalsim}
\end{equation}
where ${\cal G} \equiv  [\Z_2 \times \Z_2] \times [\Z_2 \times \Z_2] 
\simeq \Z_2^4$, with unitary representation elements corresponding to the {\em full} 
Pauli group on two qubits:
$$ \{ U_g \} = \{ \id_i , X_i , Y_i , Z_i \}^{\otimes 2} .$$
The above representation is irreducible, with $\Pi_{\cal G}$ implementing the complete depolarizing channel on two qubits: 
\[
\Pi_{\cal G}(A) = \frac{1}{16}\sum_{g\in {\cal G}} U_g^\dag A U_g = 
\frac{\tr(A)}{4}\id , \]
for every input $A$. Together with the fact that all of the system terms in $H$ are 
traceless and $F_\Gamma$ is trace-preserving, this ensures that the DD conditions 
of Eqs.~(\ref{eq:enviro_assume_3})-(\ref{eq:enviro_assume_4}) are satisfied. 
Since $|{\cal G}|=16$ and $|\Gamma|=4$, the resulting Eulerian simulation cycle will 
involve in general $N=64$ time segments, with the number of non-zero weights (hence the 
total weight $W$ and the time-overhead of the simulation) being determined by the details 
of the error model and/or the target Hamiltonian.  

A practically important case, where simpler simulation schemes are possible, occurs 
if qubits couple to their environment along a fixed axis, effectively corresponding 
to {\em pure dephasing} -- say, for concreteness, that $B_{y,i}=0=B_{z,i}$ for $i=1,2$ 
in Eq.~(\ref{openchain1}).  A smaller DD group suffices in this case \cite{KV09b}, namely 
${\cal G}_{\rm{DD}} =\Z_2$, represented again in terms of collective qubit rotations, 
\[ \{ U_h \} \equiv {G}_{\rm{D}}=
\{ \id, Z^{(\rm{all})}\} = 
\{ \id, Z_1 Z_2\}, \]
and generated by the single element $\gamma_{z,\rm{all}}$.  Clearly, the commutation 
relationship in Eq. (\ref{commute}) is maintained, still allowing our two-step procedure to be followed. 
In this case, the combined group for simulation is 
${\cal G} \equiv \Z_2 \times [\Z_2 \times \Z_2] 
\simeq \Z_2^3$, with $|{\cal G}|=8$, $|\Gamma| =3$, {\em reducibly} 
represented as follows on the two-qubit space: 
\begin{equation}
\{ U_g \} = \{ \id , X_1, Y_1, Z_1, Z_2, Z_1 Z_2, X_1 Z_2, Y_1 Z_2 \}.  
\label{dephrep}
\end{equation}

Suppose, for instance, that the task is to simulate a dipolar Hamiltonian 
$H_{\rm{dip}}$ as in Sec.~\ref{sec:example_simple}.  By following the above 
general procedure, with weights $\{ w_h\}$ given by $w_{\id} = w_{Z_1 Z_2} = 
{1}/{2}$ for $G_D$ alone, it is easy to see that Eq.~(\ref{generalsim}) simplifies, 
leading to simulation weights $ w_{\id} = {1}/{4}, w_{Z_1} = {3}/{4}= 
w_{Z_2},  w_{Z_1 Z_2} = {1}/{4},$
with the remaining 4 weights equal to 0.  While this implies that the simulation 
can now be achieved with only $N=8 \times 3 =24$ segments per cycle and minimum 
weight $W=2$, care is needed in ensuring that the DD conditions in 
Eqs. ~(\ref{eq:enviro_assume_3})-(\ref{eq:enviro_assume_4}), are {\em still} obeyed. 
This may be checked by inspection.  In particular, the fact that 
$\Pi_{\cal G}\big[F_\Gamma(X_i)\big] = 0$ for $i=1,2$ follows by analyzing the 
structure of each toggling-frame ``error Hamiltonian'', $u_{\gamma_j}^\dagger (t) X_i 
u_{\gamma_j}(t)$, for $\gamma_j \in \Gamma=\{ X_1, X_2, Z_1+Z_2\}$, 
and verifying that no term proportional to $Z_2$ is generated, that 
would be left uncorrected by averaging over the representation in Eq.~(\ref{dephrep}).
Likewise, the fact that $\Pi_{\cal G}\big[F_\Gamma(H_S)\big] = 0$ for 
$H_S=H_{\rm{iso}}$ may be directly established by a similar calculation, or by using the 
trace argument in Sec.~\ref{sec:example_simple} for the two group generators $\gamma_{x,1} =X_1$ and 
$\gamma_{z,1}=Z_1$, while also noting that for the third generator $\gamma_{z,\rm{all}}
=Z_1 Z_2$, we have $F_{Z_1 Z_2}(H_{\rm{iso}})= H_{\rm{iso}}$ and the latter is decoupled 
by the representation in Eq. (\ref{dephrep}), $\Pi_{\cal G}(H_{\rm{iso}})=0$.   
Thus, Eulerian Hamiltonian simulation in the presence of single-axis errors can be 
{\em efficiently} achieved.  
 
Again, the schemes we have just presented for $n=2$ can be generalized to 
a chain consisting of $n$ spins, which interact according to a NN Heisenberg 
interaction and are each linearly coupled to the environment, according to 
Eq.~(\ref{openchain1}).  In this case, exploiting the results of 
Sec.~\ref{sec:example_simple}, a useful group for simulation is provided by 
${\cal G}\simeq \Z_2^4$, under the unitary representation 
$$\{ U_g\} \equiv {G}_{\rm{GL}} \times {G}_{\rm{odd}}, $$
\noindent 
corresponding to generators $\gamma_{x,\rm{all}}, \gamma_{z,\rm{all}}, 
\gamma_{x,\rm{odd}}, \gamma_{z,\rm{odd}}$, all of which can be implemented 
using only 1-local (single-qubit) Hamiltonians.  As before, each simulation 
cycle will consist in the general case of arbitrary linear decoherence 
of $N=16 \times 4=64$ time segments.  
Despite the reducibility of the above representation (with the full 
Pauli group on $n$ qubits consisting of $4^n$ elements), the DD conditions 
given by Eqs.~(\ref{eq:enviro_assume_3})-(\ref{eq:enviro_assume_4}) 
remain valid for reasons similar to those outlined for $n=2$ under pure dephasing.

\subsection{Eulerian simulation of Kitaev's honeycomb lattice Hamiltonian}

\begin{figure}
\begin{center}
\hspace{5mm}\fbox{\includegraphics[width=5cm]{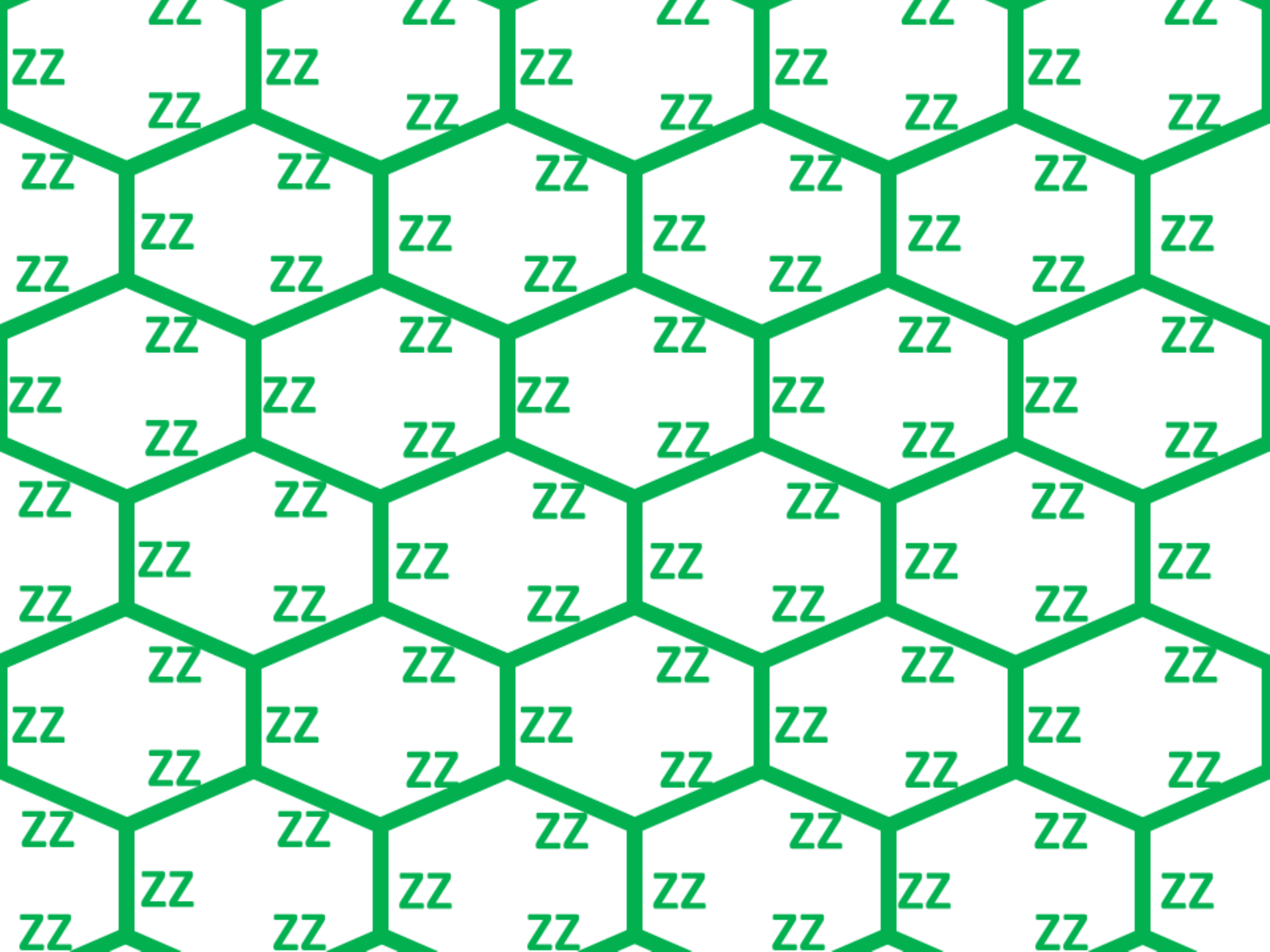}}
\hspace*{8mm}
\fbox{\includegraphics[width=5cm]{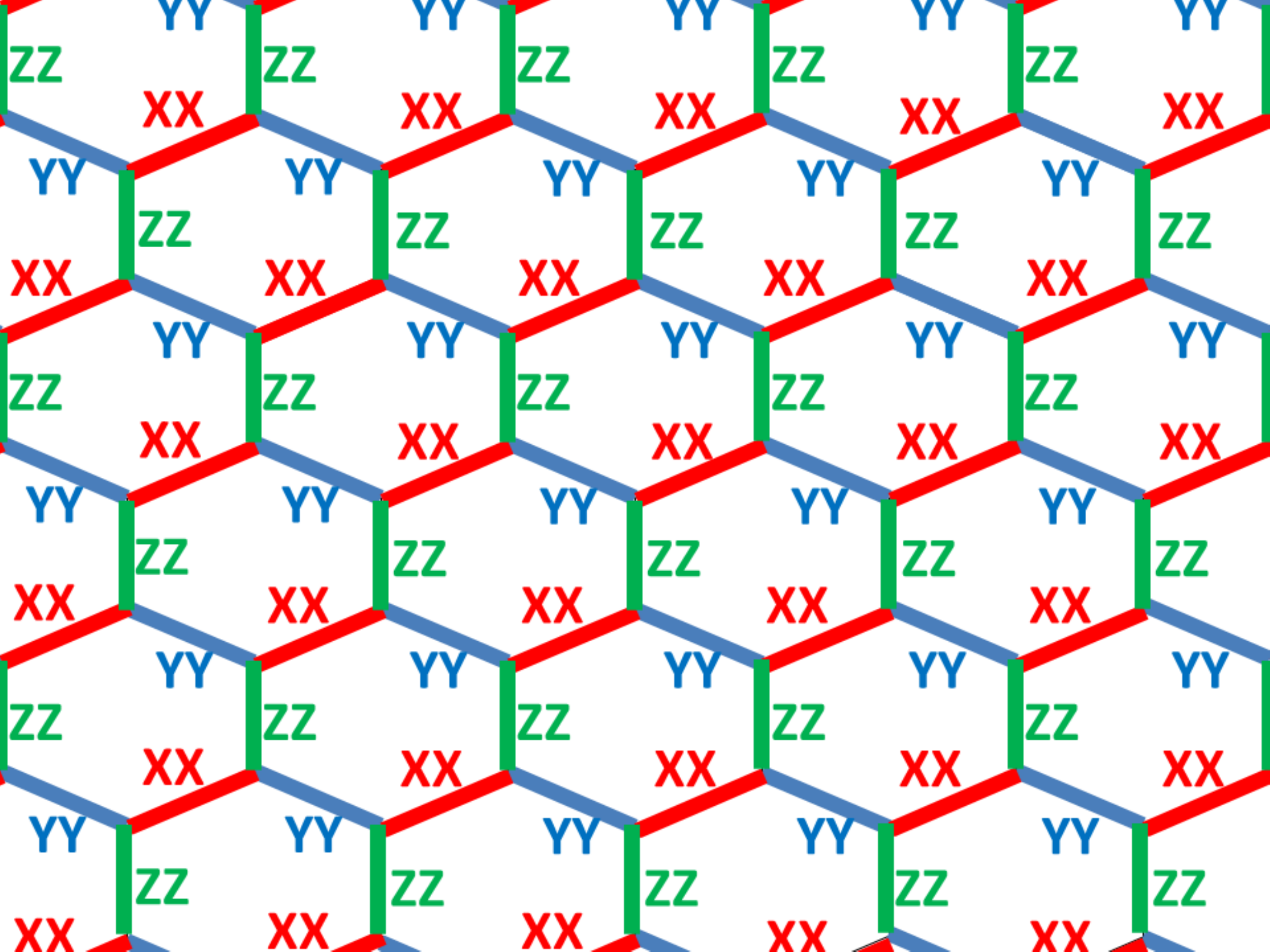}}
\end{center}
\vspace*{-2mm}
        \caption{Input and target Hamiltonians on a 2D honeycomb lattice, 
        where qubits are placed at each vertex.
        Left: The system Hamiltonian $H$ describes a system where all adjacent 
        vertices have $ZZ$ Ising couplings.
        Right: 
        The target Hamiltonian $\tilde H$ realizes Kitaev's
        honeycomb lattice model, with $XX$, $YY$, and $ZZ$ couplings
        depending on the type of the edge.}
\label{fig:honeycomb}        
\end{figure}

We return to Eulerian simulation in closed quantum systems, 
but tackle a more complicated  Hamiltonian of paradigmatic relevance to topological 
quantum memories, namely, Kitaev's honeycomb lattice model~\cite{honeycomb}. 
Suppose that the target system consists of a network of qubits arranged on 
a honeycomb lattice and interacting via NN Ising couplings.  The relevant
Hamiltonian $H$ is graphically displayed in Fig.~\ref{fig:honeycomb}(left),
where vertices represent qubits and edges represent two-qubit
couplings of the form $Z_{k} Z_{\ell}$, with vertices $k$ and
$\ell$ being adjacent in the graph and $Z_{k}$ indicating, as before, the Pauli $Z$
operator acting non-trivially only on qubit $k$.  The target Hamiltonian $\tilde H$ 
is shown in Fig.~\ref{fig:honeycomb}(right), where some of the edges are now of
the form $X_{k}X_{\ell}$ and $Y_{k} Y_{\ell}$. In accordance with the figure, 
we shall also call the $XX$-edges {\it forward-slashes}, the $YY$-edges 
{\it back-slashes}, and the $ZZ$-edges {\it verticals} henceforth. 

\begin{figure}[t]
        \centering
         \fbox{\includegraphics[width=4.9cm]{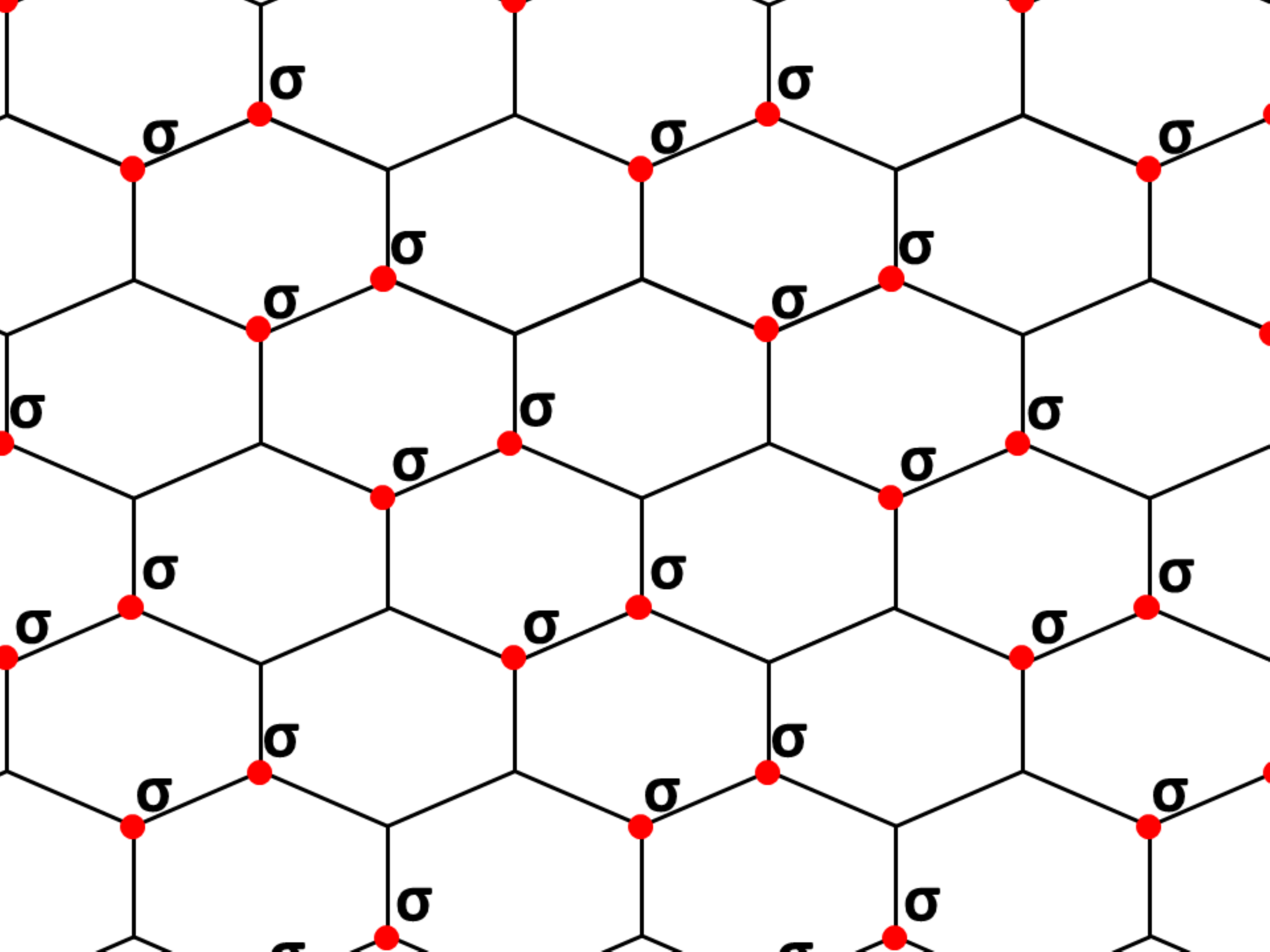}}
         \fbox{\includegraphics[width=4.9cm]{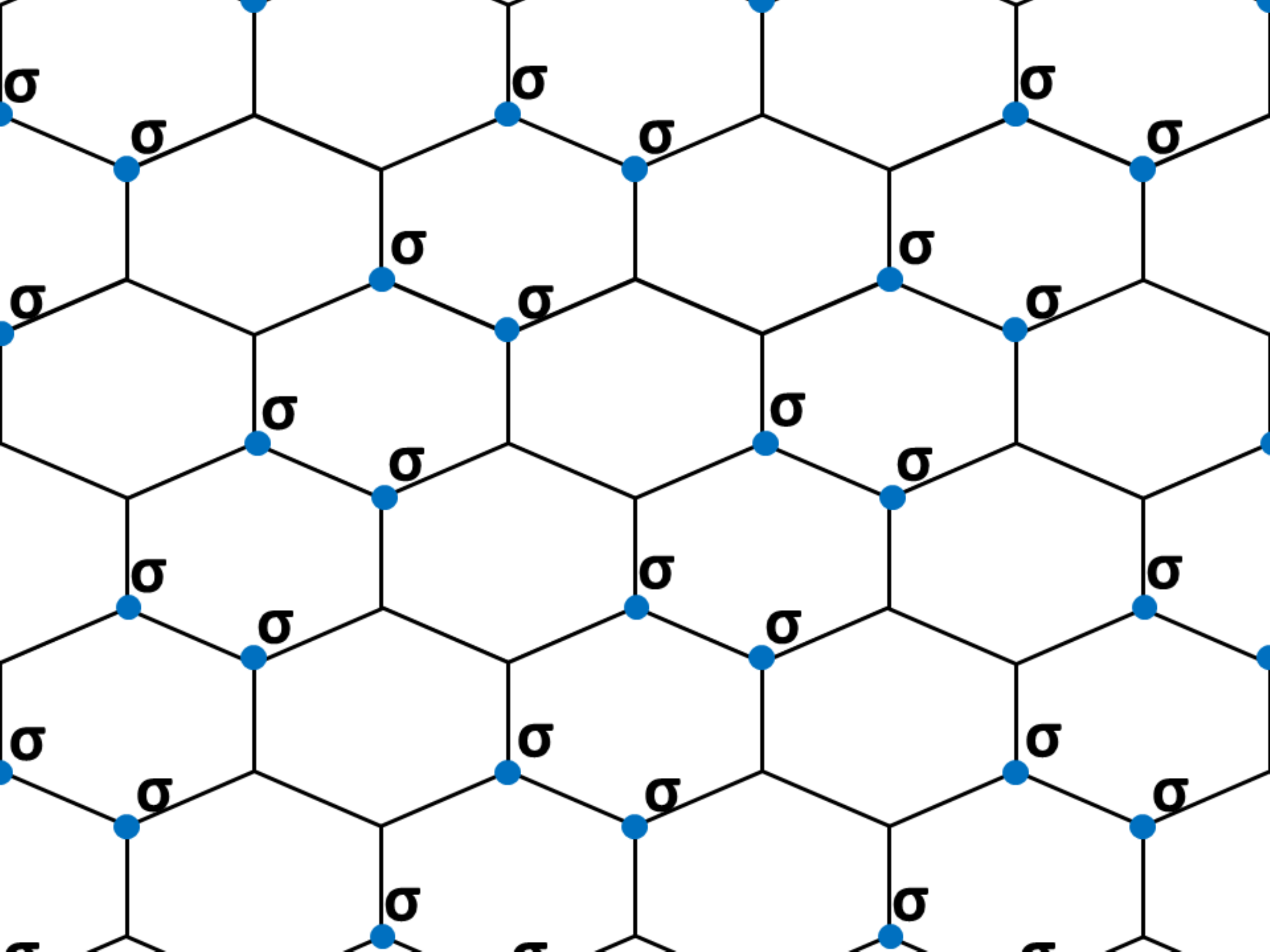}}   
         \fbox{\includegraphics[width=4.9cm]{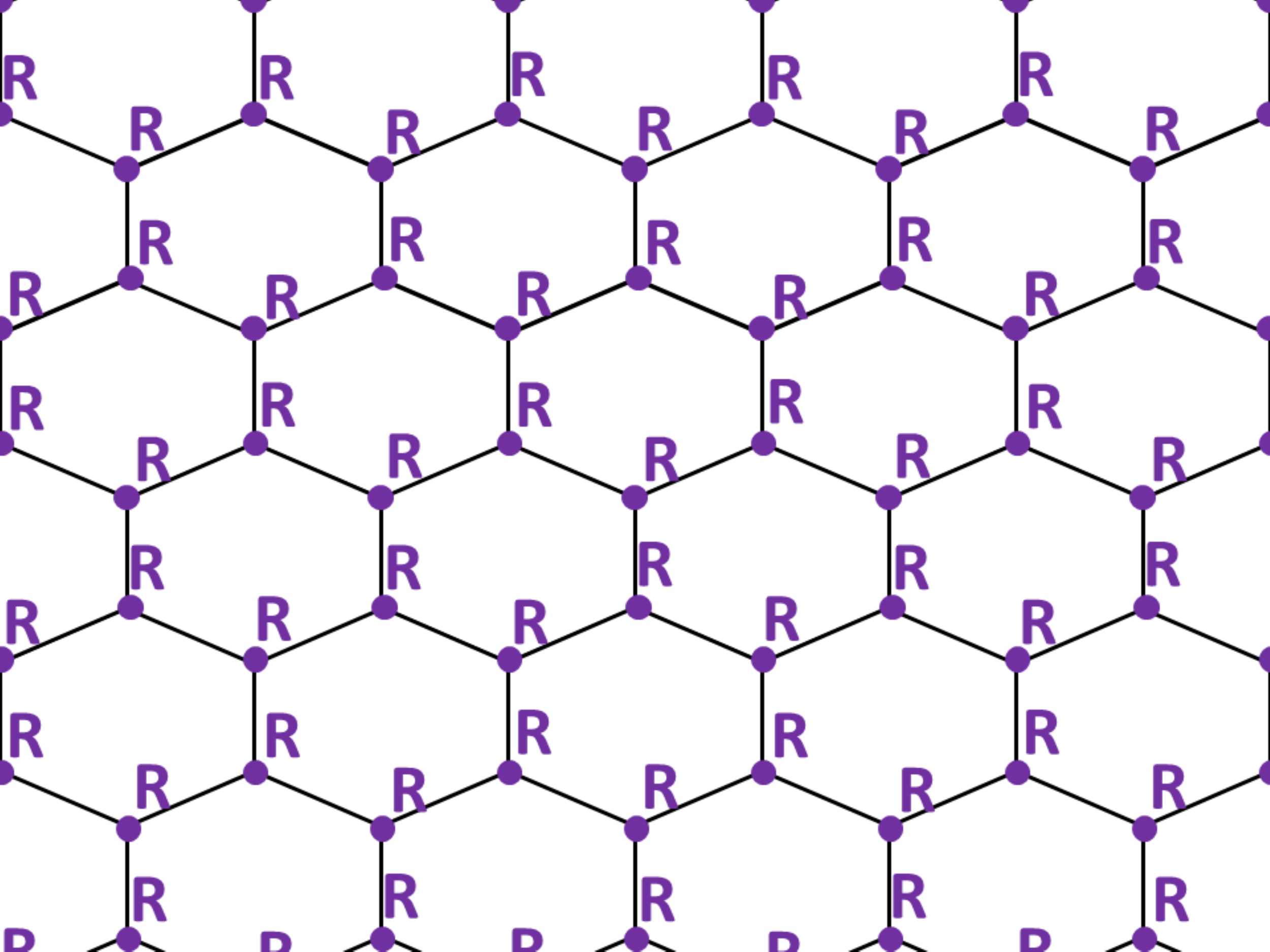}}
         \vspace*{-3mm}
\caption{Pictorial representation of different control operations. 
        Left: The unitary $\rho_\sigma$, with $\sigma$ on the
        vertices of every second forward-slash and $\id$ on all other
        vertices, where $\sigma$ is a fixed $X,Y,$ or $Z$
        operator. When $\sigma=X$, this is the generator $\rho_X$. 
        Center:
        The unitary $\tau_\sigma$, with $\sigma$ on the vertices of
        every second back-slash, where $\sigma$ is a fixed $X,Y,$ or
        $Z$ operator. When $\sigma=X$ this is the generator
        $\tau_X$. 
        Right: The generator $R_{\rm{global}}$, with $R$ at every
        vertex.}
\label{fig:group}     
\end{figure}

The basic idea to accomplish this simulation is to exploit the matrix
$R$ given in Eq.~(\ref{eq:transformerR}), in conjunction with the 
symmetry of our problem: since all Hamiltonian terms are {\em precisely} 
two-local and of the homogeneous form $\sigma\otimes\sigma$, it will be 
possible to avoid using the full machinery of a transformer.
Consider the group ${\cal G}$ generated by the three unitaries, $\rho_X,
\tau_X$, and $R_{\rm{global}}$, where $\rho_X$, shown in
Fig.~\ref{fig:group}(left) with $\sigma=X$, has $X$'s on every second
forward-slash, $\tau_X$, shown in Fig.~\ref{fig:group}(center) with
$\sigma=X$, has $X$'s on every second back-slash, and $R_{\rm{global}}$, 
shown in Fig.~\ref{fig:group}(right), has $R$ applied to every
vertex. These unitaries can be generated by one-local Hamiltonians.
By repeatedly conjugating $\rho_X$ and $\tau_X$ with $R_{\rm{global}}$, we
immediately see that we can also perform $\rho_\sigma$ and
$\tau_\sigma$, shown in Fig.~\ref{fig:group}, 
for any $\sigma = X, Y, Z$. Note that up to phase,
all such $\rho$ and $\tau$ commute. Because conjugation by $R$ maps
Pauli matrices to Pauli matrices, for any Pauli $\sigma$ we have
$R\sigma = (R\sigma R^{-1})R = \sigma' R$, where $\sigma'$ is another
Pauli matrix. Thus, up to phase, we can write any element of ${\cal G}$ 
in the canonical form
\begin{equation} 
\label{eq:honeycomb_canonical}
U_g = \rho \tau R_{\rm{global}}^a , \quad a\in\lbrace 0, 1, 2 \rbrace ,
\end{equation}
where $\rho\in\lbrace \id, \rho_X, \rho_Y, \rho_Z \rbrace, \
\tau\in\lbrace \id, \tau_X, \tau_Y, \tau_Z \rbrace,$ 
and $R^a_{\rm{global}}$ only appears on the right.

\begin{figure}
        \centering
\fbox{\includegraphics[width=4.9cm]{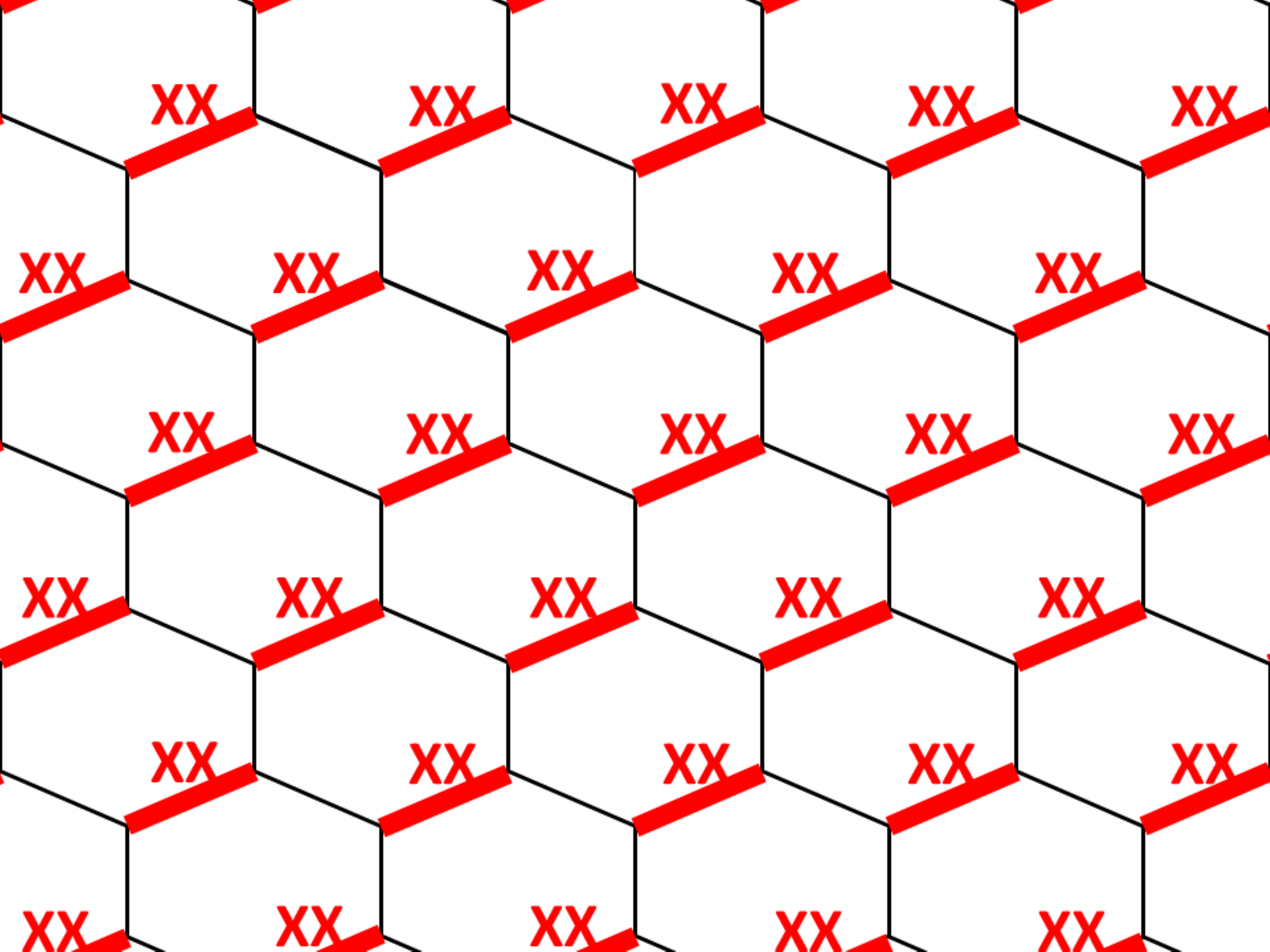}}
\fbox{\includegraphics[width=4.9cm]{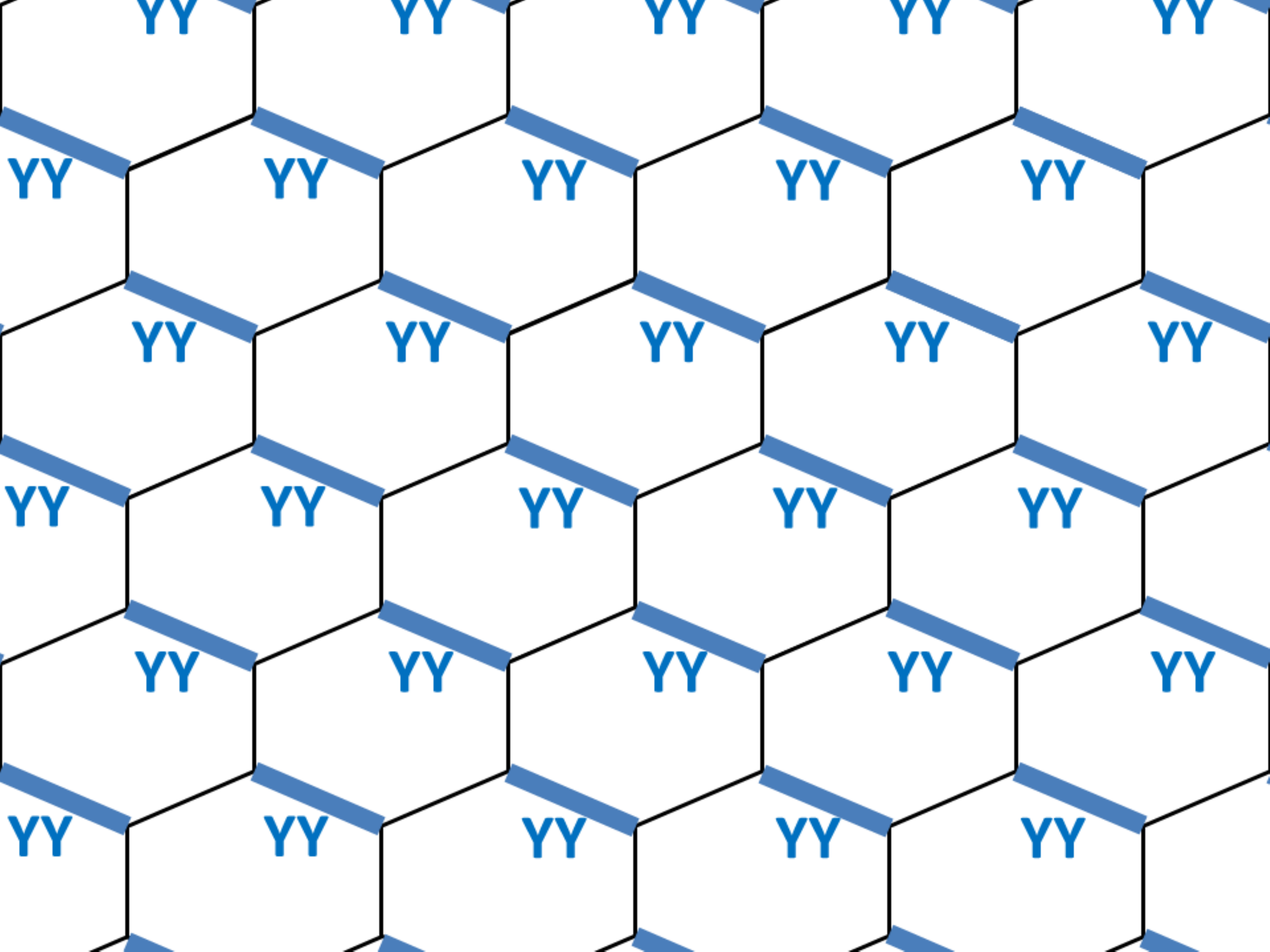}}
\fbox{\includegraphics[width=4.9cm]{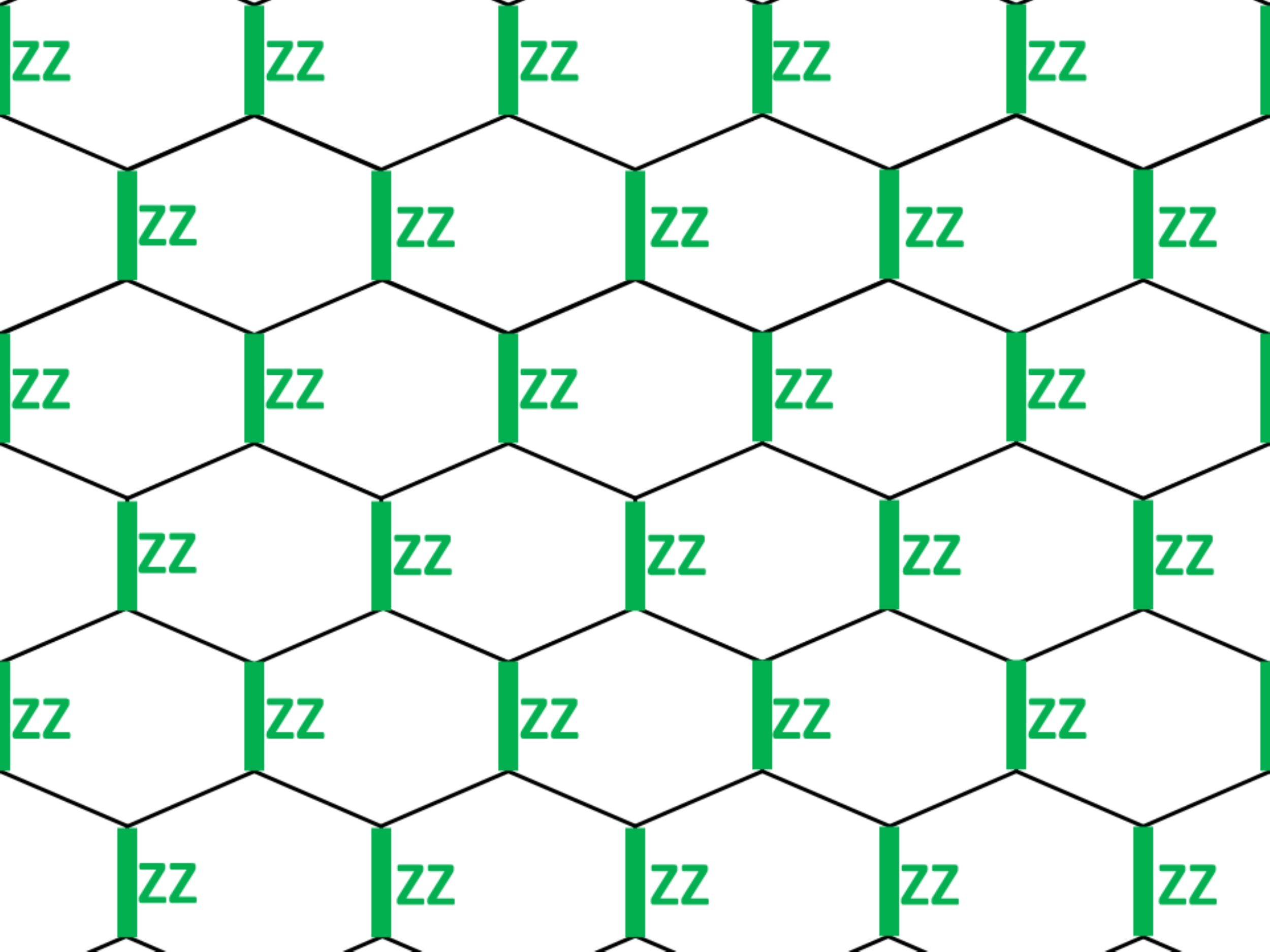}}
\vspace*{-3mm}
\caption{Pictorial representation of different simulation superoperators (see text).
        Left: Action of the superoperator $\Phi_{XX}$, 
        leaving $XX$ terms at forward-slashes only.  
        Center: Action of the superoperator $\Phi_{YY}$, 
        leaving $YY$ terms at back-slashes only. 
        Right: Action of the superoperator $\Phi_{ZZ}$,
        leaving $ZZ$ terms at verticals only.}
\label{fig:result}
\end{figure}

To construct an Eulerian simulation protocol we must be able 
to choose $w_g$ so that $\tilde{H}$ is reachable from $H$, i.e., obeys 
Eq.~(\ref{eq:bbSimulation}), while ensuring that the 
DD condition of Eq.~(\ref{eq:DDF}) is also fulfilled. We start 
from the fact that 
\[ \frac{1}{2}\id (Z\otimes Z) \id + \frac{1}{2}(X\otimes \sigma) (Z\otimes Z) (X\otimes \sigma) =
\left\{
\begin{array}{cl} 
Z\otimes Z & \rm{if}\quad \sigma=X \\
0              & \rm{if}\quad \sigma=\id
\end{array}
\right. . \]
Observe that when $U_g=\rho_X$, all forward-slash edges connect vertices that
are acted upon by either $\id\otimes\id$ or $X\otimes X$, while all
other edges connect vertices that are operated by
$X\otimes\id$. Consequently, $\frac{1}{2}\id^\dag H \id +
\frac{1}{2}\rho_X^\dag H \rho_X$ removes all Hamiltonian terms
except for those along the forward-slashes; upon conjugating by
$R_{\rm{global}}$,  we may then convert these surviving $ZZ$ terms to $XX$ terms,
as desired. To summarize,
\[ \Phi_{XX}(H) \equiv \frac{1}{2} R_{\rm{global}}^\dag H R_{\rm{global}}
+ \frac{1}{2}(\rho_X R_{\rm{global}})^\dag H (\rho_X R_{\rm{global}}) \]
gives the Hamiltonian shown in Fig.~\ref{fig:result}(left).
Similarly, the effect of $\frac{1}{2}\id^\dag H \id +
\frac{1}{2}\tau_X^\dag H \tau_X$ is to leave precisely the back-slash
edges, which can be converted from $ZZ$ to $YY$ by conjugation by
$R^2_{\rm{global}}$. Thus,
\[ \Phi_{YY}(H) \equiv \frac{1}{2} R^{2\,\dag}_{\rm{global}} H R^2_{\rm{global}} + 
\frac{1}{2}(\tau_X R^2_{\rm{global}})^\dag H (\tau_X R^2_{\rm{global}}) \]
gives the Hamiltonian shown in Fig.~\ref{fig:result}(center).
Lastly, it is not hard to see that the product $\rho_X \tau_X$ has
$X$'s on every second row of verticals; accordingly,
\[ \Phi_{ZZ}(H) \equiv \frac{1}{2}\id^\dag H \id + \frac{1}{2}(\rho_X\tau_X)^\dag H (\rho_X \tau_X) \]
isolates precisely the verticals, giving the Hamiltonian shown in
Fig.~\ref{fig:result}(right). In this case, no $R$-conjugation is
necessary since we wish to maintain $ZZ$ edges along the verticals.  
Putting all these steps together, we conclude that
\begin{eqnarray*}
\tilde H & = &
\frac{1}{2} R_{\rm{global}}^\dag H R_{\rm{global}}
+ \frac{1}{2}(\rho_X R_{\rm{global}})^\dag H (\rho_X R_{\rm{global}}) 
+ \frac{1}{2} R^{2\,\dag}_{\rm{global}} H R^2_{\rm{global}}  \\
& + &  \frac{1}{2}(\tau_X R^2_{\rm{global}})^\dag H (\tau_X R^2_{\rm{global}}) 
+ \frac{1}{2}\id^\dag H \id + \frac{1}{2}(\rho_X\tau_X)^\dag H (\rho_X \tau_X), 
\end{eqnarray*}
thus providing the desired weights for the Eulerian protocol. Since there are $|\Gamma| = 3$ 
generators and, from Eq.~(\ref{eq:honeycomb_canonical}), $|{\mathcal G}| = 4\times 4\times 3 
= 48$ group elements, each control block consists of $N =144$ time intervals. 

Lastly, we must verify that Eq.~(\ref{eq:DDF}) holds. 
Note that $F_\Gamma(H)$ acts via conjugating each vertex by unitaries (since the
generating pulses are one-local), and since such an operation is
trace-preserving at each vertex, this necessarily takes the 
precisely two-local terms in $H$ to precisely two-local terms in $F_\Gamma(H)$. 
Since no one-local terms can arise, all terms are of the form
$\sigma^{(k)}_u \otimes \sigma^{(\ell)}_v$, where $k$ and $\ell$ are adjacent
vertices and $\sigma_u, \sigma_v \in \lbrace X,Y,Z \rbrace$. Thus, we
may write
\[ F_\Gamma(H) = \sum_{k,\ell \:\rm{adjacent}} \sum_{u,v} \;a^{(k,\ell)}_{u,v}
\sigma^{(k)}_u \otimes \sigma^{(\ell)}_v \,.  \]
Due to the canonical form of our group elements, 
Eq.~(\ref{eq:honeycomb_canonical}), the action of $\Pi_{\cal G}$ reads
\[  \Pi_{\cal G}[F_\Gamma (H)] = \frac{1}{|{\cal G}|} 
\sum_{a=0}^2  \sum_{\tau , \rho } 
R^{a\dag} \tau \rho \, F_\Gamma (H) \,\rho \tau R^a ,  \]
\noindent 
where $\tau \in \{ \id, \tau_X,  \tau_Y, \tau_Z \}$  and $\rho \in \{ \id, \rho_X, \rho_Y, \rho_Z \}$, 
respectively.  Just as the map $\frac{1}{2}\id H \id + \frac{1}{2}\rho_X H \rho_X$ removes all 
non-forward-slash $ZZ$ terms, the map $\sum_{\rho} \rho F_\Gamma(H) \rho$ 
depolarizes precisely one vertex of each pair of non-forward-slash vertices, 
and therefore suppresses all non-forward-slash terms. With only
forward-slash terms remaining, $\sum_\tau \tau[ \sum_{\rho} \rho F_\Gamma(H)
\rho]\tau = 0$, since the $\tau$-sum removes all non-back-slash
terms. Thus, we conclude that $\Pi_{\cal G}\big[ F_\Gamma(H)\big] = 0$, 
as desired.

\section{Conclusion and outlook}

We have shown that the Eulerian cycle technique successfully employed in both 
dynamical decoupling schemes and dynamically corrected gates can be extended 
to also enable Hamiltonian quantum simulation with realistic {\em bounded-strength}   
controls.  For given internal dynamics and control resources, we have characterized  
the family of reachable target Hamiltonians and provided constructive open-loop 
control protocols for stroboscopically implementing a desired evolution in the 
family with accuracy (at least) up to the second order in the sense of average Hamiltonian 
theory.  We have additionally shown how Hamiltonian simulation may be accomplished 
in an open quantum system while {\em simultaneously} suppressing unwanted decoherence, 
provided that appropriate time-scale requirements and decoupling conditions 
are fulfilled.  The usefulness and flexibility of our Eulerian simulation techniques have been 
explicitly illustrated through several QIP-motivated examples involving both 
unitary and open-system dynamics on interacting qubit networks.  In all cases, 
access to purely {\em local} (single-qubit) control Hamiltonian is assumed, subject 
to finite-amplitude constraint.

It is our hope that our results may be of immediate relevance to ongoing 
efforts for developing and programming quantum simulators in the laboratory. A a 
number of possible generalizations and further theoretical questions may be worth 
considering. As an additional simulation problem dual to the one we analyzed for 
Heisenberg-coupled spin chains, exploring schemes where a target Heisenberg 
Hamiltonian is generated out of Ising couplings only would be of special interest, 
given the experimental availability of the latter in existing large-scale trapped-ion 
simulators \cite{Britton}.
Likewise, it could be useful to explore whether bounded-strength simulation 
as proposed here may be made compatible with open-loop filtering techniques for 
modulating coupling strengths, such as proposed in \cite{MikeB2013}, as well 
as in \cite{Paola2013} in conjunction with non-unitary control via field 
gradients.  Building on existing results for dynamical decoupling schemes 
\cite{LeaNJP}, the use and possible advantages of {\em randomized} simulation schemes 
in terms of efficiency and/or robustness may be yet another venue of investigation, 
especially in connection with large control groups.  Lastly, it remains an important 
open question to determine whether simulation schemes able to {\em guarantee} a 
minimum fidelity over long evolution times may be devised, in the spirit of 
\cite{Viola2013} for the particular case of the zero Hamiltonian.

\section*{Acknowledgements}
L.V. is grateful to Guifr\`e Vidal for valuable discussions and early contributions to the subject of 
this work. This research was conducted while P.W. was visiting the Center for Theoretical Physics at MIT 
during a sabbatical leave from UCF.  P.W. would like to thank Edward Farhi, Peter Shor, and their 
group members for their hospitality.  
This work was supported in part by the NSF Center for Science of Information,
under grant No. CCF-0939370, as well as by the U.S. Department of Energy under cooperative 
research agreement contract No. DE-FG02-05ER41360.
P.W. gratefully acknowledges support from NSF CAREER Award No. CCF-0746600. 
Work at Dartmouth was partially supported by the NSF under grant No. PHY-0903727, 
the U.S. Army Research Office under contract No. W911NF-11-1-0068, and the Constance and 
Walter Burke Special Project Fund in Quantum Information Science.

\section*{References}
\bibliography{Bibliography}

\begin{thebibliography}{99}
\bibitem{Remark2} Recall that a projective representation need only be a 
homomorphism up to phase, i.e., it obeys $U_{g g'} \propto U_{g} U_{g'}$ for 
$g, g' \in {\cal G}$, with proportionality rather than equality.
\bibitem{Remark3} If $\rho$ were reducible, then there would exist a
non-trivial invariant subspace $\cH_{\rm{inv}} \subset \cH_S$ such that
$\rho({\cal G}) \cH_{\rm{inv}} \subseteq \cH_{\rm{inv}}$. Hence, any 
Hamiltonian of the form $H = \sum_{ij} a_{ij} \ket{v_i}\bra{v_j}$, with $\lbrace
\ket{v_i} \rbrace$ being an orthonormal basis for $\cH_{\rm{inv}}$, could only
be transformed to other Hamiltonians of this same form, preventing 
$({\cal G},\rho)$ from being a transformer.
\end{thebibliography}

\end{document}